\documentstyle[aps,prl,preprint,floats,epsfig]{revtex}

\begin{document}

\def\gev{GeV/c$^2$}
\def\mev{MeV/c$^2$}
\def\pgev{GeV/c}
\def\km{K^-}
\def\piz{\pi^0}
\def\pip{\pi^+}
\def\ksz{\overline{K}^{*0}}
\def\dzero{$D^0$}
\def\dkppz{$D^0 \to K^- \pi^+ \pi^0$}
\def\dkppzbar{$\overline{D^0} \to K^+ \pi^- \pi^0$}

\preprint{\tighten\vbox{\hbox{\hfil CLNS 00/1700}
                        \hbox{\hfil CLEO 00-23} }}

\title{Dalitz Analysis of the Decay \dkppz}  

\author{CLEO Collaboration}
\date{\today}

\maketitle
\tighten

\begin{abstract} 
We use data collected with the CLEO II detector to perform a high-statistics 
measurement of the resonant substructure in $D^0 \to K^-\pi^+\pi^0$ decays. 
We find the Dalitz Plot is well represented by a combination of seven 
quasi-two-body decay channels ($\ksz \piz$, $K^- \rho$, $K^{*-} \pip$, $K_0(1430)^-\pip$, 
$\overline{K}_0(1430)^0 \piz$, $K^- \rho^+(1700)$, and $K^*(1680)^- \pip$), 
plus a small non-resonant component.  
We see no evidence of a scalar $\kappa\rightarrow K^-\pi^+$ resonance 
in the mass range recently reported by other groups.
Using the amplitudes and phases from 
this analysis, we calculate an integrated CP asymmetry of $-0.031 \pm 0.086$.
\end{abstract}
\newpage

\begin{center}
S.~Kopp,$^{1}$ M.~Kostin,$^{1}$
A.~H.~Mahmood,$^{2}$
S.~E.~Csorna,$^{3}$ I.~Danko,$^{3}$ K.~W.~McLean,$^{3}$
Z.~Xu,$^{3}$
R.~Godang,$^{4}$
G.~Bonvicini,$^{5}$ D.~Cinabro,$^{5}$
% ...
\end{center}
 
\small
\begin{center}
$^{1}${University of Texas, Austin, TX  78712}\\
$^{2}${University of Texas - Pan American, Edinburg, TX 78539}\\
$^{3}${Vanderbilt University, Nashville, Tennessee 37235}\\
$^{4}${Virginia Polytechnic Institute and State University,
Blacksburg, Virginia 24061}\\
$^{5}${Wayne State University, Detroit, Michigan 48202}\\
% ...
\end{center}

\newpage

\section{Introduction}

A clearer understanding of final state interactions in exclusive weak decays 
is an important ingredient for our ability to model 
decay rates as well as for our understanding of interesting phenomena 
such as mixing~\cite{ref:hepph9802291}.
There are several theoretical methods 
~\cite{ref:bsw,ref:bedaque,ref:chau,ref:terasaki,ref:buccella} 
used to understand the dynamics of two body charmed meson decays
with experimental measurements as input.  Unfortunately, final-state 
interactions are often not well understood, and may not be
included properly in many models. These
long-distance strong interaction effects can cause significant changes in decay
rates for specific final states, and can cause shifts in the phases of the
decay amplitudes.

Three-body decays provide a rich laboratory in which to study the
interference between intermediate state resonances, and provide a 
direct probe of the final state interactions in certain decays.
When a particle decays into three or more 
daughters, intermediate resonances dominate the decay rate. These
resonances will cause a non-uniform distribution of events in phase space
when analyzed using a ``Dalitz Plot'' technique~\cite{ref:dalitz}. 
Since all events of a particular decay mode have 
the same final state, multiple resonances at the same
location in phase space will interfere. This provides the opportunity
to experimentally measure both the amplitudes and phases of the 
intermediate decay channels, which in turn allows us to deduce their
relative branching fractions.  These phase differences can even allow 
details about very broad resonances to be extracted by observing 
their interference with other intermediate states.

This paper describes a study of the underlying structure in
\dkppz ~decays~\cite{ref:cc}. Four previous groups have made 
similar measurements, each with
less than 1000 signal events\cite{ref:e687,ref:e691,ref:mk31,ref:e516}.
In addition to significantly increasing our statistical power, the large 
CLEO II dataset on which this analysis is based also permits us to 
tighten analysis requirements to drastically reduce the effect of backgrounds.
Coupled with the superb resolution of the CLEO-II detector, 
this has allowed us to extract significantly more information
about this decay than was possible in the past.

\section{Theoretical Models}
\label{sec:theory}
Since we are studying the decay of a spin-zero particle to three daughters,
only two degrees-of-freedom are required to completely describe the
kinematics. To see this, consider the
decay in the \dzero ~rest frame. The four-momenta of the three final
state particles correspond to twelve unknowns. We have one
constraint for each known mass and four additional constraints from
the conservation of momentum and energy in the decay.  Finally, 
since the three degrees of freedom describing the spatial orientation of 
the decay are irrelevant (the \dzero ~having spin zero) only two independent 
degrees of freedom remain. 

There are three invariant masses that can be formed by considering
all possible pairs of final state particles: 
$M^2_{K^-\pi^+}$, $M^2_{K^-\pi^0}$ and $M^2_{\pi^+\pi^0}$.  
Only two of these are independent, however, since
energy and momentum conservation results in the additional constraint
\begin{equation}
\label{eq:epcons}
M^2_{D^0} + M^2_{K^-} + M^2_{\pi^+} + M^2_{\pi^0} = M^2_{K^-\pi^+}
+ M^2_{K^-\pi^0} + M^2_{\pi^+\pi^0}.
\end{equation}

Choosing two of the above three invariant masses as dynamic variables
has two compelling advantages: i) Their relativistic invariance means the Lorentz 
frame in which they are evaluated is irrelevant; ii) As the expression for the
partial width in Equation~\ref{eq:dgamma} indicates,
we expect that a scatter plot of events in the $M_{12}^2$ vs $M_{23}^2$ plane 
(known as a Dalitz Plot) will be uniformly distributed
if phase-space alone determines the decay dynamics.  This allows the 
decay fraction at each point to be readily
correlated with the decay matrix element $\cal M$ without additional
corrections:

\begin{equation}
\label{eq:dgamma}
d\Gamma = { |{\cal M}|^2 \over{256 \pi^3 M_D^3}} dM_{12}^2
dM_{23}^2.
\end{equation}

\begin{figure}
\centerline{\psfig{file=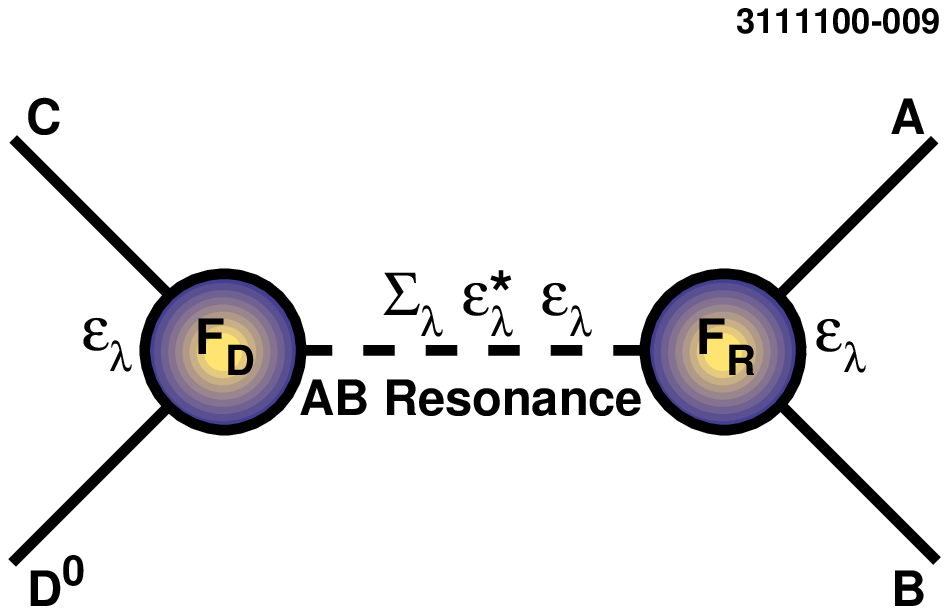,width=6in}}
\vspace{0.5cm}
\caption{\label{fig:fyn}A Representation of the three body decay
of $D^0 \to A B C$ through an AB resonance. The spin sum
is performed to obtain the angular dependence of the decay.}
\end{figure}

In this analysis we choose $M^2_{K^-\pi^+}$ and $M^2_{\pi^+\pi^0}$ as our 
two Dalitz Plot variables, and must next construct 
the relevant decay amplitudes in terms of these. 
The dynamics can be understood with the aid of
Figure~\ref{fig:fyn}. We first consider the decay of the $D^0$ meson
into particle C plus an AB resonance, followed by the decay of the AB
resonance into particles A and B. To properly describe the structure
of this decay using our Dalitz Plot variables, we need to obtain the
angular dependence of the decay products. Each vertex in
Figure~\ref{fig:fyn} contains a spin factor $\varepsilon_\lambda$ which
depends on the type of the decay: scalar, vector, tensor, etc. 
The matrix element for a vector decay is
\begin{equation} 
\label{eq:vecspin} 
{\cal M} = F_D (P_{D^0} + P_C)_\mu {\sum_\lambda
\varepsilon^{\mu*}_\lambda \varepsilon^\nu_\lambda \over{M^2_r - M^2_{AB} -
i M_r \Gamma_{AB}}} (P_A - P_B)_\nu F_r 
\end{equation}
where $P$ denotes 4-momentum, and $M_r$ is the mass of the resonance. 
In general the form factors at each vertex, $F_D$ and $F_r$, are unknown functions,
however in practice they are either set to a constant value or
to the Blatt-Weisskopf penetration factors~\cite{ref:blatt}.

The spin-sum in the numerator of Equation~\ref{eq:vecspin} is
evaluated to give
\begin{equation}
\label{eq:spinsum}
\sum_\lambda \varepsilon^{\mu*}_\lambda \varepsilon^\nu_\lambda = 
 -g^{\mu\nu} + {{P_{AB}^\mu P_{AB}^\nu}\over{M_{AB}^2}}
\end{equation} 
and the ``mass dependent width'' $\Gamma_{AB}$ is a function of 
the AB invariant mass $M_{AB}$, 
the momentum of either daughter in the AB rest frame $p_{AB}$, 
the momentum of either daughter in the resonance rest frame $p_r$, 
the spin of the resonance $J$, 
the width of the resonance $\Gamma_r$, and is expressed as~\cite{ref:pilkuhn}:
\begin{equation}
\Gamma_{AB} = \Gamma_r \left( p_{AB} \over{p_r} \right)^{2J+1} \left( M_r \over{M_{AB}} \right) F_r^2
\end{equation} 

We relax the transversality requirement on the vector resonance in 
Equation~\ref{eq:spinsum} and divide by
$M_r^2$ instead of $M_{AB}^2$. This substitution
gives rise to a small spin zero component when the vector resonance is
off mass-shell, a behavior which is observed to occur with the $W$ boson
and which should also be expected in the resonance behavior we are studying
here.  

Inserting this expression for the spin-sum into
Equation~\ref{eq:vecspin} and summing over the repeated indices gives
the Lorentz invariant expression for the matrix element of a vector
particle as a function of position in the Dalitz Plot:
\begin{equation} 
\label{eq:svec}
{\cal A}_1(ABC|r) = F_D F_r {M^2_{AC} - M^2_{BC} + { (M^2_D - M^2_C)(M^2_B -
M^2_A)\over{M^2_{r}}} \over {M^2_r - M^2_{AB} - i M_r \Gamma_{AB}}}.
\end{equation}

The procedure for calculating the vector matrix element is 
generalizable to intermediate particles having other spin. For example, 
we can easily find the amplitude for a spin zero resonance to be
\begin{equation}
\label{eq:sscalar}
{\cal A}_0(ABC|r) = F_D F_r { 1 \over {M^2_r - M^2_{AB} - i M_r \Gamma_{AB}}}.
\end{equation}

The procedure for higher spin resonances involves a bit more algebra. 
For example, the spin two case starts with 
\begin{equation}
\label{eq:mten}
{\cal A}_2(ABC|r) = F_D (P_D + P_C)_\mu (P_D + P_C)_\nu {\sum_\lambda
\varepsilon^{\mu\nu*}_\lambda \varepsilon^{\alpha\beta}_\lambda \over{M^2_r - M^2_{AB} -
i M_r \Gamma_{AB}}} (P_A - P_B)_\alpha (P_A - P_B)_\beta  F_r. 
\end{equation} 
In this case the spin sum has been previously calculated by Pilkuhn~\cite{ref:pilkuhn} to be
\begin{equation}
\sum\limits_\lambda  {\varepsilon _\lambda ^{*\mu \nu } \varepsilon _\lambda ^{\alpha \beta } }  
= \frac{1}{2}\left( {T ^{\mu \alpha } T ^{\nu \beta }  + 
T ^{\mu \beta } T ^{\nu \alpha } } \right) - 
\frac{1}{3}T ^{\mu \nu } T ^{\alpha \beta } 
\end{equation}
where 
\begin{equation}
T^{\mu \nu} = -g^{\mu\nu} + {{P^\mu P^\nu}\over{M^2}}
\end{equation}

When this expression is inserted into
Equation~\ref{eq:mten} and the implied sums performed 
we find the final form of the tensor matrix element:
$$
{\cal A}_2(ABC|r)  =  {F_D F_r \over{M_r^2 - M^2_{AB} - i\Gamma_{AB} M_r}} 
\left[ \left( M^2_{BC} - M^2_{AC} + {(M_D^2 - M_C^2) (M_A^2 - M_B^2) 
\over{M_{r}^2}}\right)^2 \right .
$$
\begin{equation}
-\frac{1}{3} \left(M^2_{AB} - 2 M_D^2 - 2 M_C^2 + {( M_D^2 - M_C^2)^2 
 \over{ M_{r}^2}}\right)\left. \left(M^2_{AB} - 2 M_A^2 - 2 M_B^2 + 
 {( M^2_A - M^2_B)^2 \over{ M_{r}^2}}\right) \right].
\end{equation}

Next we return to the form factors $F_D$ and $F_r$, which attempt to
model the underlying quark structure of the $D^0$ meson and 
the intermediate resonances.  We use the Blatt-Weisskopf penetration factors
shown in Table~\ref{tbl:bwff}. These have one free parameter, R, which
is the ``radius'' of the meson, and are dependent on the momentum $P$ of the
decay particles in the parent rest frame. In all cases, we normalize
the form factor to have unit value at the nominal meson mass. The fits
display very little sensitivity to the meson radii; good fits
are obtained when these values vary between $0\ {\rm GeV}^{-1}$ and
$10\ {\rm GeV}^{-1}$ for the $D^0$ and between $0\ {\rm GeV}^{-1}$ and
$3\ {\rm GeV}^{-1}$ for the intermediate resonances. To be consistent with
other experiments~\cite{ref:e687} we have chosen the $D^0$ to have 
$R=5\ {\rm GeV}^{-1}$ and the intermediate resonances all to have 
$R=1.5\ {\rm GeV}^{-1}$

\begin{table}
\begin{center}
\caption{\label{tbl:bwff} Blatt-Weisskopf Penetration Form Factors. 
$p_r$ is the momentum of either daughter in the meson rest frame.
$p_{AB}$ is the momentum of either daughter in the candidate rest frame (same
as $p_r$ except the parent mass used is the two-track invariant mass of the candidate
rather than the mass of the meson). $R$ is the meson radial parameter.}
\vspace{0.5cm}
\begin{tabular}{|c|c|} 
Spin & Form Factor $ F_r $ \\ \hline
0 & 1 \\
& \\
1 & ${\sqrt{1 + R^2 p^2_r}\over{\sqrt{1+R^2 p^2_{AB}}}} $ \\
& \\
2 & ${\sqrt{9 + 3 R^2 p^2_r + R^4 p^4_r }\over{\sqrt{9+ 3 R^2 p^2_{AB} + R^4 p^4_{AB}}}} $ \\ 
\end{tabular}
\end{center}
\end{table}

Before continuing, we must specify our phase conventions for the
intermediate resonances.  We can explicitly see the importance of
specifying the ordering of particles in the decay by examining
Equation~\ref{eq:svec}. If we were to switch the labels A and B we
would generate an overall minus sign causing the phase to
change by 180$^o$. In an attempt to be consistent with previous results 
we have chosen the phases in the same way as the E687 collaboration~\cite{ref:e687} 
since they are the only group to have explicitly published their choice of phases
and matrix elements. 

\medskip
Now that we know the form of the intermediate resonance amplitudes, 
and have chosen a phase convention that will allow us to compare 
our results with previous measurements, we can write down an expression
for the overall matrix element of the decay. Guided by the results of previous
measurements~\cite{ref:e687,ref:e691,ref:mk31}, 
we begin with only three vector resonances $\rho(770)^+$, $K^{*-}$
and $\bar{K}^{*0}$~\cite{ref:kstar} as well as a flat non-resonant ({\it nr}) component:
\begin{eqnarray}
\label{eq:decm}
{\cal M}(D^0 \to K^-\pi^+\pi^0) & = & a_{nr} e^{i\phi_{nr}} \nonumber \\
&& + a_{\rho} e^{i\phi_{\rho}} {\cal A}_1(\pi^+\pi^0K^-|\rho^+) \nonumber \\
&& + a_{\bar{K}^{*0}} e^{i\phi_{\bar{K}^{*0}}} {\cal A}_1(K^-\pi^+\pi^0|\bar{K}^{*0}) \nonumber \\
&& + a_{K^{*-}} e^{i\phi_{K^{*-}}} {\cal A}_1(K^-\pi^0\pi^+|K^{*-}),
\end{eqnarray}
where the $a_i$ and $\phi_i$ are the amplitude and relative phase of
the $i$'th component respectively. 
The overall normalization is arbitrary, and is chosen to be
\begin{equation}
\label{eq:dpint}
 \int | {\cal M} |^2 d{\cal DP} = 1 
\end{equation}
where $d{\cal DP}$ indicates that the integral is performed over the Dalitz Plot.
This is equivalent to saying that we are sensitive only to relative phases 
and amplitudes, which in turn means that we are free to fix one phase and one
amplitude in Equation~\ref{eq:decm}. To minimize 
correlated errors on the phases and amplitudes we choose the largest mode, 
$K^-\rho$, to have a fixed zero phase and an amplitude of one.

Since the choice of normalization, phase convention, and amplitude
formalism may not always be identical for different experiments, 
fit fractions are reported instead of amplitudes to allow for more meaningful 
comparisons between results. The fit fraction is defined as the integral of a single
component divided by the coherent sum of all components:
\begin{equation}
\label{eq:ddpint}
{\rm Fit\ Fraction} =  {\int \left| a_r e^{i\phi_r} {\cal A}(A B C |r) \right |^2 d{\cal DP} \over
{ \int \left| \sum_j  a_j e^{i\phi_j} {\cal A}(A B C |j) \right |^2 d{\cal DP}} }.
\end{equation}
 
The sum of the fit fractions for all components of a fit
will in general not be one because of interference.

One must also consider the statistical errors on the fit fractions. We have
chosen to use the full covariance matrix from the fits to determine
the errors on fit fractions so that the assigned errors will properly
include the correlated components of the errors on the amplitudes and
phases. After each fit, the covariance matrix and final parameter
values are used to generate 500 sample parameter sets.  For each set,
the fit fractions are calculated and recorded in histograms. Each
histogram is fit with a single Gaussian to extract its width, which
is used as a measure of the statistical error on the fit fraction.

\section{Experimental Details}
\label{sec:dalcuts}
The CLEO II detector is described elsewhere~\cite{ref:cleoii}.
This measurement uses the entire CLEO II dataset,
which represents approximately $4.7$ fb$^{-1}$ of integrated
 $e^+e^-$ luminosity at $\sqrt{s}\sim 10.6$ GeV. 

The $D^0$'s used in this analysis are required to be produced by the decay
chain $D^{*+} \to D^0 \pi^+_s$, which
significantly reduces the combinatorial background. 
To reconstruct the $D^0$'s, we take pairs of
oppositely charged tracks and assign the track with the same sign as
the pion from the $D^{*+}$ decay to be the pion from the $D^0$ decay. This
Cabibbo-favored correlation between the signs of the pions eliminates 
the need for other particle identification techniques in this analysis.
  
For tracks to be used they must be well fitted, reconstruct to
within 5 cm of the interaction point along the beam pipe and within 5
mm perpendicular to the beam pipe (corresponding to about 5 standard 
deviations in length and more than 10 standard deviations in the 
width of the beam spot).  

We fit pairs of
tracks passing these requirements to a common vertex, which is the
candidate decay position of the $D^0$ meson.  Each such pair of
charged tracks is combined with all $\piz$ candidates in an event.
The $\piz$ candidates are found by combining all pairs of
electromagnetic showers which are unmatched to charged tracks. 
To reduce the number of fake $\piz$'s from random shower
combinations and to improve their resolution, we require that each
shower have energy above 100 MeV and be in the central region of the CLEO II
detector. Furthermore, the invariant mass of the two photon
combination is restricted to be between 128~\mev and 140~\mev
({\it i.e.} within about one standard deviation of the $\piz$ mass). 
The two shower combination is kinematically fit to give the known $\piz$ mass.

Once we have a vertex with a $\km$, a $\pip$ and a $\piz$
candidate, we combine the momenta of the three particles to find the
$D^0$ momentum. With the decay location and the momentum known, the
$D^0$ is projected back to the beam spot. 
In CLEO, the beam spot has a ribbon like
shape with a width of 700 $\mu$m, a height of 20 $\mu$m, and a length of 
about 2 cm.  We project the
$D^0$ candidate back to the vertical position of the beam, since this
dimension of the beam is most precisely known. 
The intersection of the $D^0$ projection and the beam position 
defines the production point of the $D^{*+}$.

We refit the slow pion track to  
include the $D^{*+}$ production point as an additional constraint,
providing a better measurement of its true momentum. 
The result of this is that the width of the mass difference peak, 
$\Delta M =M(D^{*+}) - M(D^0)$, is reduced from 590~keV to
490~keV, providing a 15\% reduction in the number of background events
in our final sample. We make a requirement
that $\Delta M$ is between 144.9~\mev and 145.9~\mev. We also
require that the 
normalized $D^*$ momentum, $X_{D^*} ={ P_{D^*}/\sqrt{E^2_{beam} -
M^2_{D^*}}} $, is greater than 0.6, which significantly reduces the 
combinatorial background level and kinematically excludes 
the possibility that a $D^*$ candidate came from a decaying $B$ meson.

\begin{figure} 
\centerline{\psfig{file=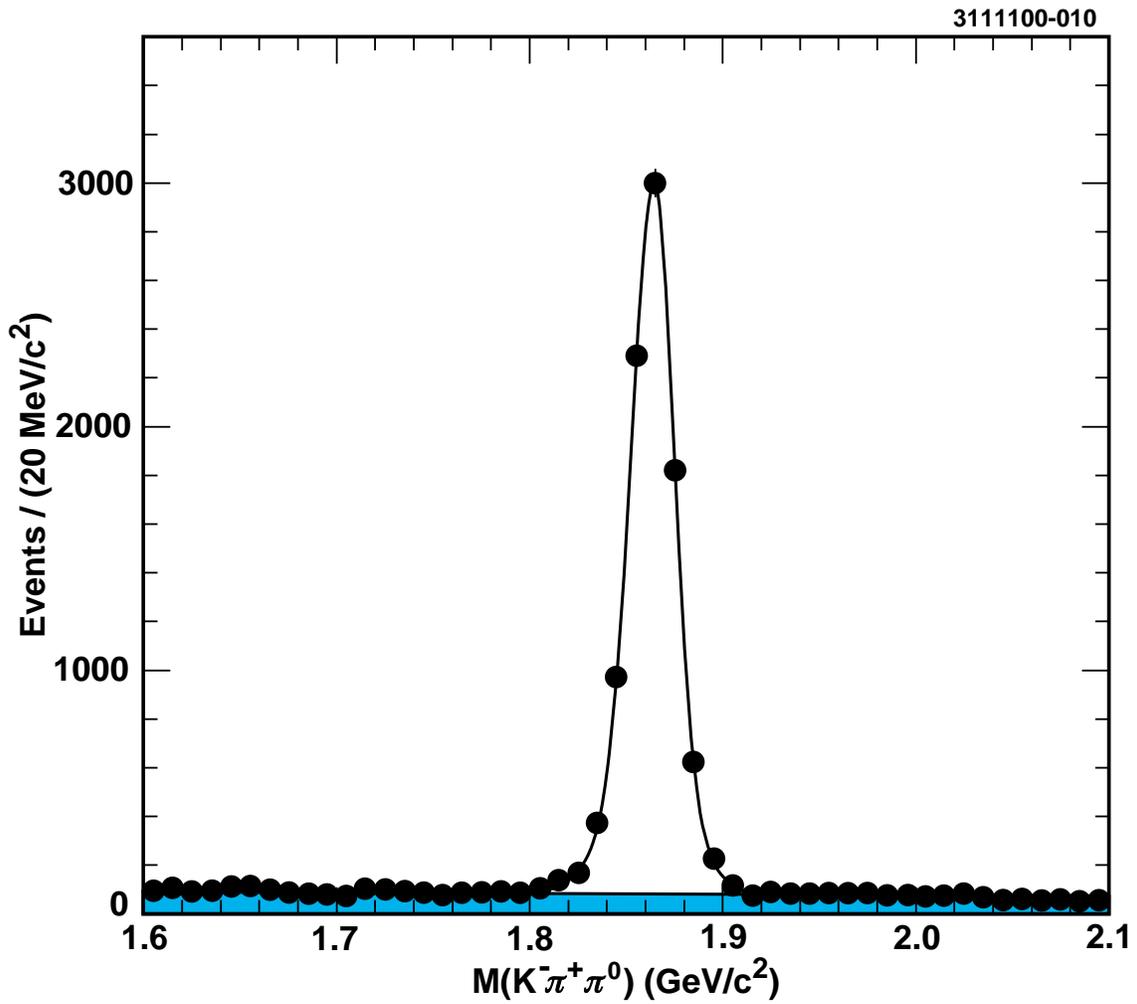,width=6in}}
\vspace{0.5cm}
\caption{The \dkppz ~reconstructed mass distribution for all event candidates (points).
The solid line represents a fit to the data using a double bifurcated Gaussian to
represent the signal plus a straight line to represent the background (shaded).}
\label{fig:md0}
\end{figure}

After obtaining the candidate $D^0$'s as described above, we can plot
the mass of the \dkppz ~candidates as shown in Figure~\ref{fig:md0},
where the fit to the mass distribution is also shown.
When examining the Dalitz Plot, we only use the events which have
$1.85\, {\rm GeV/c}^2 < M_{D^0} < 1.88\, {\rm GeV/c}^2$ ({\it i.e.} within about 
one standard deviation of the known $D^0$ mass).

We have chosen quite restrictive cuts on our kinematic variables (approximately 
one standard deviation on each) to minimize the effect of the background on our
result. Since we are studying the shape of the distribution and are not trying
to extract a branching ratio, the fact that this increases the systematic 
uncertainty of the overall efficiency somewhat is not an issue.  

\begin{figure} 
\centerline{\psfig{file=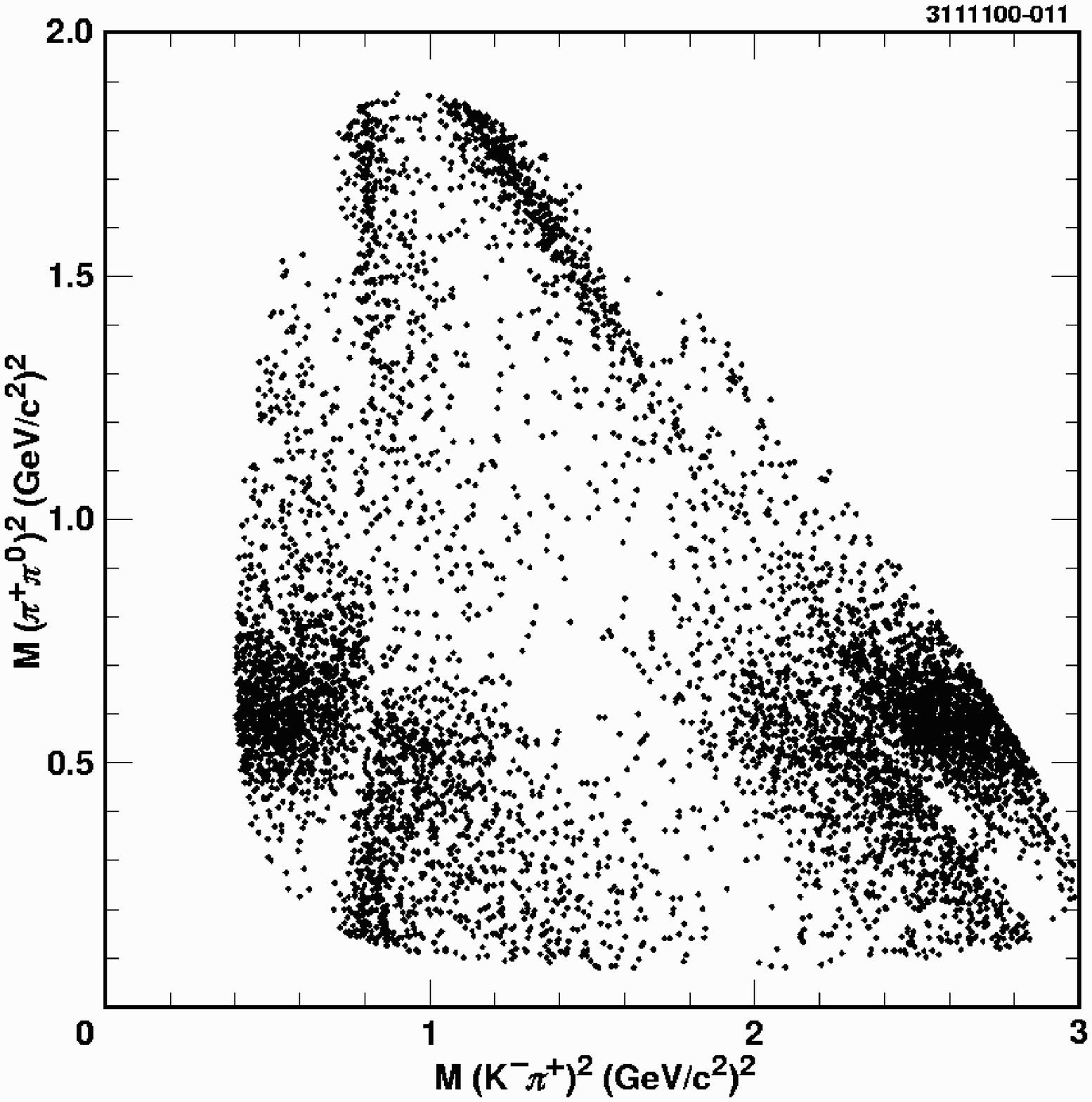,width=4in}}
\vspace{0.5cm}
\caption{\label{fig:daldata}The Dalitz distribution of all 7,070 \dkppz ~candidates in 
our data sample shown in an unbinned scatter plot.}
\end{figure}

Applying the above requirements produces 7,070 events in the Dalitz Plot.  
Figure~\ref{fig:daldata} shows the distribution of this sample
as a scatter plot in the chosen mass squared variables
$M_{K^-\pip}^2$ and $M_{\pip\piz}^2$.

In order to reduce the smearing effects introduced by the detector,
those combinations passing the above requirements are kinematically
fit such that when combined, the $\km$, $\pip$ and $\piz$ reconstruct
to give the correct $D^0$ mass. This kinematic fit has two
effects.  First, the uncertainty of the 4-momentum of the particles is
reduced, giving a more precise measurement of the mass squared
variables used to define an event's position in the Dalitz Plot.
Second, the decay position in these variables is guaranteed to
respect the kinematic boundaries of the Dalitz Plot. 

\subsection{Background}

Turning again to Figure~\ref{fig:md0}, we can see that the signal
region contains a small but non-zero
number of background events. We use the fit shown
to measure the fraction of events in this region which are
``true signal'' by integrating the signal function (a double
bifurcated Gaussian) and the background function (a line) and
comparing the two. The signal fraction and its associated statistical 
error, $0.967 \pm 0.011$, are used in the likelihood function 
minimized during the fitting procedure.

Knowing only the amount of background is not enough if we want to
correctly extract the amplitudes and phases of the signal component;
the shape of the background in the Dalitz Plot is also important. 
There are several sidebands~\cite{sideband} that could be chosen to study the shape of 
this background using data, and a Monte Carlo study (outlined below) 
is used to determine which one is best. As will become apparent in the 
section on systematic errors, the overall low level of the background 
means that the final result has very little sensitivity to this
choice.

To determine which sideband will best represent the Dalitz Plot 
shape of the background in the signal region, a signal-free sample of
 $e^+e^-\rightarrow q\overline{q}$ Monte Carlo simulated data 
is used (referred to below as the ``vetoed'' sample).  
These data are generated using a full GEANT based detector simulation~\cite{ref:geant}, 
and are processed by the same reconstruction code that is used for real data.

This Monte Carlo sample represents the background we
want to measure in data, and we use it as a reference in our study. The next step
is to consider a number of possible sideband samples, and see which
 does the best job representing the Dalitz shape of the vetoed sample.
 
Many sideband samples can be formed in the space defined by the three mass
variables, $\Delta M$, $M_{D^0}$ and $M_{\pi^0}$.
To choose the best one, we fit the distribution in the
Dalitz Plot using an unbinned likelihood fit to a cubic polynomial in
$M^2_{K^-\pi^+}$ and $M^2_{\pi^+\pi^0}$ as well as non-interfering
squared amplitudes for the $\rho(770)$, $K^*(892)^-$ and $\ksz$. 
A $\chi^2$ is formed between each Monte Carlo sideband sample and the 
reference vetoed sample to give us a measure of their relative merits.

\begin{figure} 
\centerline{\psfig{file=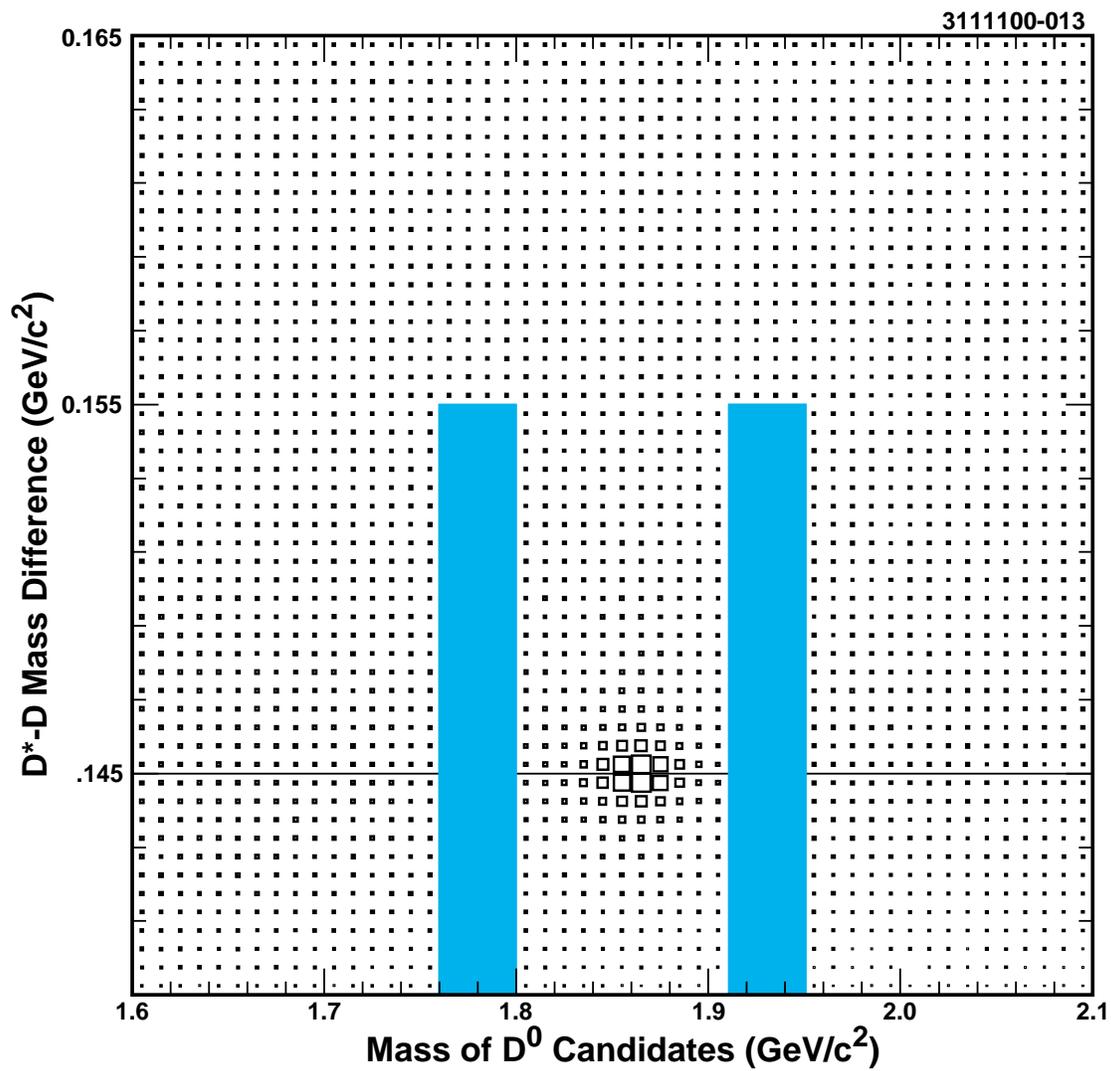,width=6in}}
\vspace{0.5cm}
\caption{\label{fig:2dside} A box plot showing \dkppz
~candidates in the $\Delta M$ vs $M_{D^0}$ plane.
The shaded areas on each side of the peak are the sidebands determined to
best represent the background.}
\end{figure}

Based on this, the sideband which seems to best represent the vetoed 
sample consists of those events which have $\Delta M < 0.1549$ \gev, 
are in the $M_{\pi^0}$ signal region, and are in
the off-peak regions of $M_{D^0}$: $1.76 < M_{D^0}{\rm (GeV/c^2)} < 1.80$ or
$1.91 < M_{D^0}{\rm (GeV/c^2)} < 1.95$. This 
choice of sidebands along with the \dkppz ~candidates are shown 
in Figure~\ref{fig:2dside}.  

\begin{figure} 
\centerline{\psfig{file=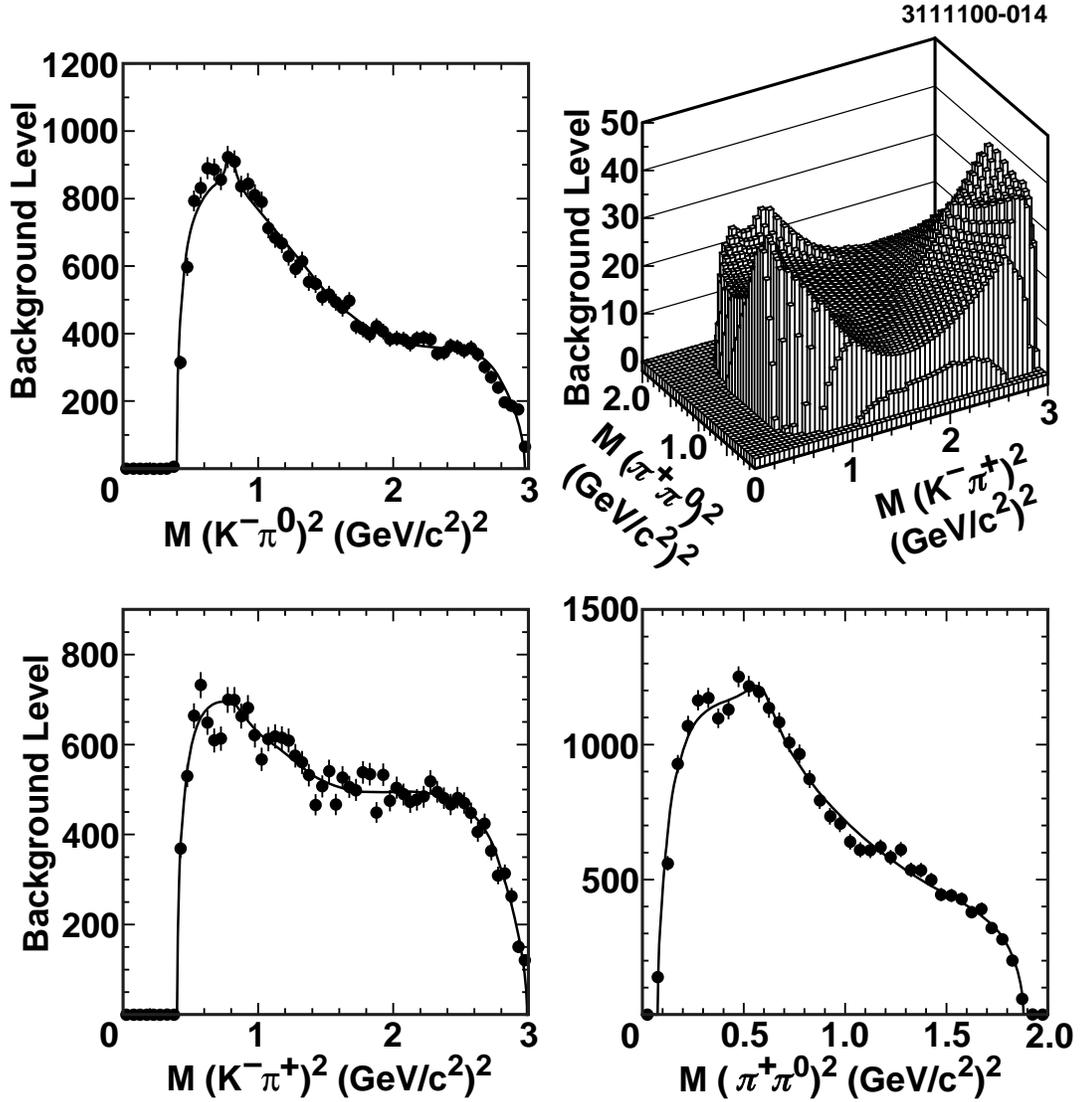,width=6in}}
\vspace{0.5cm}
\caption{\label{fig:back}
Results of the best fit to the \dkppz ~sideband background sample.
The fit function is shown as a Manhattan plot (top right), and as projections onto
the three mass squared variables showing both fit (histogram) and data (points).}
\end{figure}

The assumption is now made that the sideband method which best represents 
the background in the Dalitz Plot when analyzing the Monte Carlo simulated
data is also the best sideband method for use in real data.
Those events from the actual data which are in the selected ``best'' 
sideband are fit
with the cubic polynomial plus the three non-interfering resonances.
The resulting best fit parameters are shown in
Table~\ref{tbl:backeff}.  We project the fit and the background points
onto the three mass squared variables and show the results in
Figure~\ref{fig:back}, along with a two dimensional Manhattan plot
of the fit result. We use this parameterization of the background
shape in the fit to the distribution of events in the Dalitz Plot by
including both the parameters and the covariance matrix in the final
likelihood function (as described in Section~\ref{sec:fittest}).

\begin{table} 
\begin{center}
\caption{\label{tbl:backeff}Background and efficiency best fit parameters.
The fitting functions are described in Section~\ref{sec:fittest}.}
\vspace{0.5cm}
\begin{tabular}{|c|c|c|c|} 
\multicolumn{2}{|c|}{Background} & \multicolumn{2}{c|}{Efficiency} \\
\hline 
$B_0$        & $ 1.0 $ (fixed)       &$E_0$      & $ (22.1 \pm 1.8)\times 10^{-5}$ \\
$B_x$        & $ -1.188 \pm 0.018 $  &$E_x$      & $ (-6.89 \pm 2.9)\times 10^{-5}$\\
$B_y$        & $ -0.742 \pm 0.044 $  &$E_y$      & $ (-27.1 \pm 3.7)\times 10^{-5}$\\
$B_{x^2}$    & $0.483 \pm 0.015 $    &$E_{x^2}$  & $ (10.4 \pm 1.6)\times 10^{-5}$\\
$B_{xy} $    & $ 0.874 \pm 0.032 $   &$E_{xy} $  & $  (38.2 \pm 3.2)\times 10^{-5}$\\
$B_{y^2}$    & $ 0.122 \pm  0.047 $  &$E_{y^2}$  & $ (12.4 \pm 2.8)\times 10^{-5}$\\
$B_{x^3}$    & $ -0.052 \pm  0.004 $ &$E_{x^3}$  & $ (-3.00 \pm 0.27)\times 10^{-5}$\\
$B_{x^2y}$   & $ -0.162 \pm  0.010 $ &$E_{x^2y}$ & $ (-7.97 \pm 0.73)\times 10^{-5}$\\
$B_{xy^2}$   & $ -0.202 \pm 0.014 $  &$E_{xy^2}$ & $ (-12.8 \pm 0.94)\times 10^{-5}$\\
$B_{y^3}$    & $0.061 \pm  0.016 $   &$E_{y^3}$  & $ (-0.53 \pm 0.73)\times 10^{-5}$\\ 
$B_{\ksz}$   & $(1.65 \pm  1.70) \times 10^{-5} $ && \\
$B_{\rho^+}$ & $(3.69 \pm 0.58) \times 10^{-4} $ && \\ 
$B_{K^{*-}}$ & $(7.69 \pm 1.95) \times 10^{-5} $ && \\ 
\end{tabular} 
\end{center} 
\end{table} 

\subsection{Efficiency}

Next, we determine the efficiency for detecting signal events as a function
of position in the two dimensional Dalitz Plot. After generating
4.2~million signal Monte Carlo events with a flat distribution in
phase space ({\it i.e.}, uniform across the Dalitz Plot), these events are
analyzed to find the number of observed events as a function of
$M^2_{K^-\pi^+}$ and $M^2_{\pi^+\pi^0}$. The events observed are binned into
regions with $50~({\rm MeV/c}^2)^2$ on a side, and we divide the
number of events observed in each bin by the number generated to give
a measure of the efficiency for that bin. Due to the finite number of
Monte Carlo events observed in each bin, each individual efficiency
measurement has about a 10\% statistical error. Since we expect (and 
observe) that the
efficiency is a slowly varying function across the Dalitz Plot, we
fit the efficiency measurements with a cubic polynomial in
$M^2_{K^-\pi^+}$ and $M^2_{\pi^+\pi^0}$ and use the resulting function
to parameterize the efficiency.

\begin{figure} 
\centerline{\psfig{file=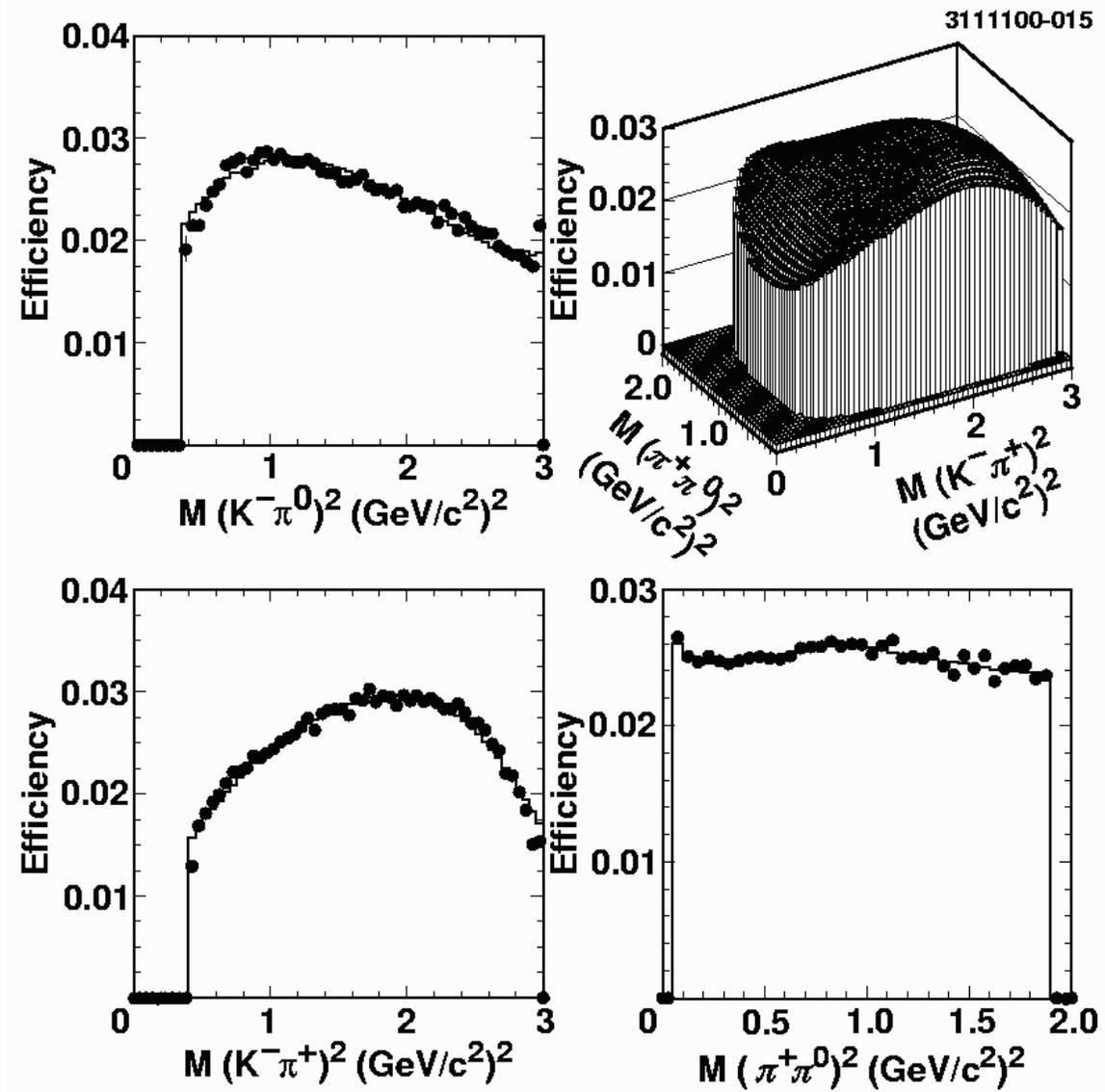,width=6in}}
\vspace{0.5cm}
\caption{\label{fig:eff} 
Results of the best fit to the \dkppz ~detection efficiency distribution.
The fit function is shown as a Manhattan plot (top right), and as projections onto
the three mass squared variables of both fit (histogram) and raw efficiency (points).
In each of the projections, the quantity plotted is the average efficiency at the given
$M^2$ value.}
\end{figure}

As a check that the efficiency function obtained using phase space
Monte Carlo is reasonable, we repeat the procedure described above with
another 2.4~million signal Monte Carlo events generated with the
Dalitz distribution found by E691~\cite{ref:e691}. Again the
efficiency in each bin is calculated and fit.  Since the resulting fit
agrees well with the efficiency calculated from the phase space
distribution of points, we combine the two Monte Carlo samples and
calculate the efficiency using the full 6.6~million events.  The fit
parameters for this combined fit are shown in
Table~\ref{tbl:backeff}. Figure~\ref{fig:eff} shows the raw efficiency
for each bin as well as the fit and the projections onto each of the 
three mass-squared axes. 

\section{Fitting Procedure}
\label{sec:fittest}
Having a parameterization for both the background and efficiency as
well as knowing the fraction of events in the signal region 
which are in fact background, we can fit the
data in the Dalitz Plot to extract the amplitudes and phases of any
contributing intermediate resonances. 

To do this we use an unbinned maximum likelihood fit which minimizes the
function ${\cal F}$ given by
\begin{equation} 
{\cal F} = {\big[\sum_{events}-2\ln {\cal L}\big] + \chi^2_{\rm penalty} }
\end{equation} 
where
\begin{equation} 
\label{eq:likelihood}
{\cal L} =  \left({F{{\cal E}(M^2_{K^-\pi^+}, M^2_{\pi^+\pi^0}) 
\left| {\cal M} \right|^2 }\over{ {\cal N}_{signal}}} 
+ (1 - F) {{{\cal B}(M^2_{K^-\pi^+}, M^2_{\pi^+\pi^0})}\over{{\cal N}_{background}}}\right ) 
\end{equation}
\begin{eqnarray} 
\label{eq:chisquared}
\chi^2_{\rm penalty} & =  \left( {F - F_o \over{\sigma_F}} \right)^2 + \sum_{ij} 
(B_i-B_{io}) V_{ij} (B_j-B_{jo})  \nonumber \\
&+ E_{sys}\sum_{ij} (E_i-E_{io}) W_{ij} (E_j-E_{jo}) 
\end{eqnarray}
and 
\begin{equation}
{\cal N}_{signal} = \int {{\cal E}(M^2_{K^-\pi^+}, M^2_{\pi^+\pi^0})
\left| {\cal M} \right|^2} d{\cal DP}
\end{equation}
\begin{equation}
{\cal N}_{background} = \int {\cal B}(M^2_{K^-\pi^+}, M^2_{\pi^+\pi^0})
d{\cal DP}.
\end{equation}

The signal fraction $F_o$ and its error $\sigma_F$ (0.967 and 0.011 respectively) 
are determined 
from the fit to the \dzero ~mass spectrum shown in Figure~\ref{fig:md0}, and the
the parameters $B_{jo}$ and $E_{jo}$ describe the nominal background
and efficiency shapes (see Table~\ref{tbl:backeff}) via the cubic polynomial
shapes
\begin{eqnarray}
 {\cal B} & = & B_0 + B_x M^2_{K^-\pi^+} + B_y M^2_{\pi^+\pi^0} + 
B_{x^2} (M^2_{K^-\pi^+})^2 + B_{xy} M^2_{K^-\pi^+} M^2_{\pi^+\pi^0} + 
B_{y^2} (M^2_{\pi^+\pi^0})^2 + \nonumber \\
& & B_{x^3} (M^2_{K^-\pi^+})^3 + 
B_{x^2y} (M^2_{K^-\pi^+})^2 M^2_{\pi^+\pi^0} + 
B_{xy^2} M^2_{K^-\pi^+} (M^2_{\pi^+\pi^0})^2 + 
B_{y^3} (M^2_{\pi^+\pi^0})^3 +\nonumber \\
& & B_{\ksz} | {\cal A}_1(\km \pip \piz | \ksz )|^2 +
\beta_{\rho} |{\cal A}_1(\pip \piz \km | \rho^+)|^2 +
\beta_{K^{*-}} |{\cal A}_1(\km \piz \pip | K^{*-})|^2 
\end{eqnarray}
and
\begin{eqnarray}
 {\cal E} & = & E_0 + E_x M^2_{K^-\pi^+} + E_y M^2_{\pi^+\pi^0} + 
E_{x^2} (M^2_{K^-\pi^+})^2 + E_{xy} M^2_{K^-\pi^+} M^2_{\pi^+\pi^0} + 
E_{y^2} (M^2_{\pi^+\pi^0})^2 + \nonumber \\
& & E_{x^3} (M^2_{K^-\pi^+})^3 + 
E_{x^2y} (M^2_{K^-\pi^+})^2 M^2_{\pi^+\pi^0} + 
E_{xy^2} M^2_{K^-\pi^+} (M^2_{\pi^+\pi^0})^2 + E_{y^3} (M^2_{\pi^+\pi^0})^3 
\end{eqnarray}.

In expressing this likelihood function we have made the explicit
assumption that background events and signal events are distinct,
allowing us to factor the likelihood into two components
which do not interfere. The $\chi^2_{\rm penalty}$ terms 
represent the information from the fits used to determine
the signal fraction, the background parameterization, and the 
efficiency parameterization. $V_{ij}$ and $W_{ij}$ are the 
covariance matrices from the background and efficiency fits respectively. 
The last term is used only when evaluating the systematic errors due to the 
efficiency parameterization, hence $E_{sys}$ is set to zero during ``normal'' 
fitting. 

In addition to the likelihood, we need a measure to assess 
how well any given fit represents the data.  A
confidence level can be calculated directly
from the likelihood function by utilizing the best fit
parameters. This idea was described by ARGUS~\cite{ref:goodfit}
and is a direct application of the Central Limit Theorem from
statistics~\cite{ref:clt}. 
Assuming the candidates are truly distributed
according to the likelihood function which gives the best fit, the
average value is
\begin{equation}
\mu = \frac{1}{N}\sum_{i=1}^N ( -2 \ln {\cal L}) \approx 
\int {\cal L} ( -2 \ln {\cal L}) d{\cal DP}
\end{equation}
where $N$ is the number of candidates.  The variance is given by 
\begin{equation}
\sigma^2_\mu =  \frac{1}{N}\sum_{i=1}^N ( -2 \ln {\cal L} - \mu)^2 \approx
\int {\cal L} ( -2 \ln {\cal L})^2 d{\cal DP} - \mu^2.
\end{equation}

Because we have a large number of candidates distributed according to this
function, the Central Limit Theorem tells us that the mean should
follow a normal distribution. The sum of minus log likelihoods, which is the
value minimized in the fit, has a mean of $N\mu$ and follows a normal
distribution with a variance of $N\sigma^2_\mu$. Thus, the minimal
value will come from a normal distribution with mean
\begin{equation}
<-2 \sum \ln {\cal L}> = N \int {\cal L} ( -2 \ln {\cal L}) d{\cal DP} - n
\end{equation}
and standard deviation
\begin{equation}
 \sigma_{<-2 \sum \ln {\cal L}>} = \sqrt{N \int {\cal L} 
 ( -2 \ln {\cal L})^2 d{\cal DP} - N\mu^2 }
\end{equation}
where $n$ is the number of parameters extracted from the fit. 
The confidence level for the fit is then just the area of a Gaussian with
the above mean and width which lies above the value obtained in our fit.
It is worth pointing out that this value only gives a measurement
of the goodness of fit assuming the fit function correctly describes
the true distribution. 

Having a second measure of the goodness of the fit 
would be extremely valuable, and an obvious choice is the $\chi^2$.
This requires the data to be binned, and furthermore that there are 
enough events in each bin that Gaussian statistics can be assumed. 
As we saw in Figure~\ref{fig:daldata}, the density of candidates in the \dkppz 
~Dalitz Plot varies significantly as a function of position, hence to form a 
sensible $\chi^2$ measure we will need to have bins of varying size.

To systematically choose these bins, we start by placing a grid 
of small regions, $50~({\rm MeV/c}^2)^2$ on a side, over
the Dalitz Plot. Next, adjacent regions are combined into bins until each 
contains approximately 30 candidates.  
After completing this procedure, our Dalitz Plot is divided into 228 bins of
varying size, and a $\chi^2$ variable for the
multinomial distribution~\cite{ref:lnchi2,ref:eadie} can be calculated as
\begin{equation}
\chi^2 = -2 \sum_{i=1}^{228} n_i \ln
\left(\frac{p_i}{n_i}\right)
\end{equation}
where $n_i$ is the number of events observed in bin $i$, and $p_i$
is the number predicted from the fit.
For a large number of events this formulation of the $\chi^2$ 
becomes equivalent to the usual one~\cite{ref:dof}.  

One can naively calculate the number of degrees of freedom
for the fit as the number of bins (r) minus the number of fit parameters (k)
minus one, as would be correct for a binned maximum likelihood
fit. However, since we are minimizing the unbinned likelihood function,
our ``$\chi^2$'' variable does not asymptotically follow a $\chi^2$
distribution~\cite{ref:dof}, but it is bounded by a $\chi^2$ variable
with $(r-1)$ degrees of freedom and a $\chi^2$ variable with $(r-k-1)$
degrees of freedom. Because it is bounded by two $\chi^2$ variables, it 
should be a useful statistic for comparing the relative goodness of fits.
In what follows, we use both the $\chi^2$ and the confidence level described above 
as our ``goodness of fit'' measures to determine which of the many 
possible sets of intermediate resonances are preferred.  

Before analyzing 
the \dkppz ~data, we performed many checks of both the fitting and 
fit evaluation procedures.  One of these was a double-blind study in which several
Monte Carlo samples containing \dkppz ~decays generated with ``secret'' mixtures
of intermediate resonances were analyzed. In each case, our fitting and evaluation 
procedure identified the correct set of resonances, and recovered their amplitudes 
and phases within statistical errors. The resulting amplitudes and phases for 
one of the fits is shown in Table~\ref{tbl:ajwfit}.

\begin{table}
\begin{center}
\caption{\label{tbl:ajwfit} A comparison between input Monte Carlo parameters 
and the results from a subsequent fit to the Dalitz Plot using the techniques
described in Section~\ref{sec:fittest}. 
Note that the input amplitudes and phases are completely fictitious.}
\vspace{0.2cm}
\begin{tabular}{|l|cc|cc|}
Resonance & \multicolumn{2}{c|}{Generated} & \multicolumn{2}{c|}{Measured} \\ 
& Amplitude & Phase (degrees)& Amplitude & Phase (degrees) \\ \hline
$\ksz$       &  1.0 & 45  & $1.03 \pm 0.02$ & $47 \pm 1 $ \\
$\rho^+$     &  1.0 &  0  & $1.0$ (fixed) & $ 0$ (fixed) \\
$K^{*-}$     &  1.0 & -115& $1.03 \pm 0.02$ & $-113 \pm 2 $ \\
$K^*_0(1430)^-$&  0.5 & -115& $0.54 \pm 0.05$ & $-107 \pm 6 $ \\
Non resonant &  1.0 &  -90& $1.08 \pm 0.05$ & $273 \pm 3 $ \\ 
\end{tabular}
\end{center}
\end{table}

\section{Fitting the Data}
\label{sec:datafits}

Armed with the tools described in the previous section, we are ready to fit the
data distribution shown in Figure~\ref{fig:daldata}. Previous
experiments have observed three intermediate resonances in \dkppz ~decays: $\rho^+$,
$\ksz$ and $K^{*-}$, hence we begin by considering only these in
addition to a non-resonant component. The resulting fit parameters are
given in Table~\ref{tbl:fit}.

\begin{table}
\begin{center}
\caption{\label{tbl:fit} Results of the best fit to the data with only $\rho^+$,
$\ksz$, $K^{*-}$, and non-resonant components included.}
\vspace{0.2cm}
\begin{tabular}{|lcr|lcr|} 
\hspace{0.2in}
&$a_{nr}$ &\hspace{0.2in}&\hspace{0.2in}&  $ 1.70 \pm 0.07 $&\hspace{0.2in} \\
&$a_{\rho^+}$ &&& $ 1.00 $ (fixed)& \\
&$a_{K^{*-}}$ &&& $ 0.378 \pm 0.008 $& \\
&$a_{\overline{K}^{*0}}$ &&& $0.422 \pm 0.009 $&\\ \hline
&$\phi_{nr}$ &&& $ 59.7^o \pm 2.0^o $ &\\ 
&$\phi_{\rho^+}$ &&& $ 0^o $ (fixed)& \\
&$\phi_{K^{*-}}$&&& $ 166.7 \pm 2.0^o$& \\
&$\phi_{\overline{K}^{*0}}$ &&& $ -7.8^o \pm 2.2^o$&\\ \hline
&$-2 \ln {\cal L}$ &&& 7070 & \\
&Conf. Level. &&& 0.0\% & \\
&$\chi^2$ &&& 650 & \\ 
\end{tabular}
\end{center}
\end{table} 

\begin{figure} 
\centerline{\psfig{file=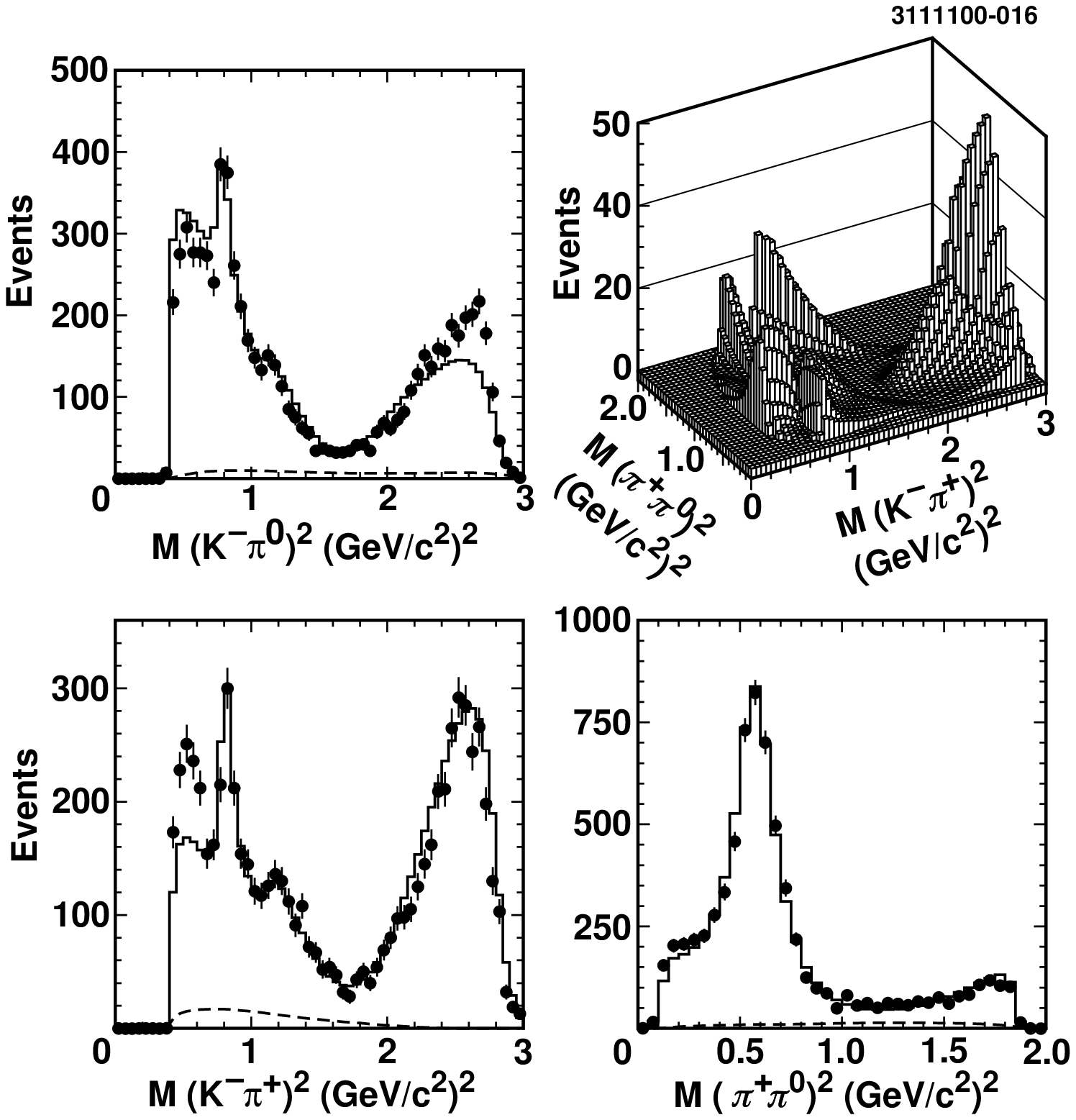,width=6in}}
\vspace{0.5cm}
\caption{\label{fig:s3fit} The results of fitting the
\dkppz ~data with only $\rho^+$, $\ksz$, $K^{*-}$, 
and non resonant components included.
The efficiency corrected fit is shown as a Manhattan plot (top right), 
and as projections onto the three mass squared variables of both 
fit (histogram) and data (points).
The dashed line shows the level of the background.}
\end{figure}

Figure~\ref{fig:s3fit} shows the projections of both the fit and the
data onto the three mass squared variables, as well as a two
dimensional Manhattan plot of the final fit function. Even a quick glance
suggests that the data is not well represented by this function, 
and the large value of 
$\chi^2$ as well as the zero confidence level confirm this
observation. These parameters are useful for comparison with
previous experiments, however, which reported observation of these
three resonances with much less statistics.  We show the comparison in
Tables~\ref{tbl:resultsff} and~\ref{tbl:resultsphase} and see good
agreement. Unfortunately, we can only compare the results for the
phases to E687 since the other experiments do not give their choice of
particle ordering or potential complex constants in their choice for
${\cal A}(ABC|r)$. Although the phases match for the three resonant
components, the non-resonant phase seems to be off by 180$^o$.
This observation is consistent with comments that E687 had an unreported
negative sign in their vector amplitude~\cite{ref:jew}.

\begin{table}
\caption{\label{tbl:resultsff}A comparison of the fit fractions obtained with 
our ``three resonance'' fit and those reported by previous experiments. The
errors shown are statistical only. Note that although the data is not well fit by this
model, the results are consistent with those reported by previous experiments.}
\vspace{0.2cm}
\begin{tabular}{ccccc}
Decay Mode & CLEO II (3 Resonance)&E687 & Mark III & E691 \\ \hline
$K^-\rho^+$   & $0.834 \pm 0.007$ & $0.765 \pm 0.041 $ &$0.81 \pm 0.03 $& $0.647 \pm 0.039$   \\
$K^{*-}\pi^+$ & $0.129 \pm 0.006$  & $0.148 \pm 0.028 $ &$0.12 \pm 0.02 $ & $0.084 \pm 0.011$    \\
$\overline{K}^{*0}\pi^0$& $0.157 \pm 0.007 $  & $0.165 \pm 0.031 $ &$0.13 \pm 0.02$& $0.142\pm 0.018$ \\
Non resonant   & $0.074\pm 0.006 $  & $0.101 \pm 0.033 $ &$0.09 \pm 0.02 $&$0.036$ \\
\end{tabular}
\end{table} 

\begin{table}
\caption{\label{tbl:resultsphase}A comparison of the phases (in degrees) obtained with 
our ``three resonance'' fit and those reported by previous experiments. The
errors shown are statistical only. 
In the ``Rotated'' column we have shifted the reported phases such that the $\rho$ has 
a phase of $0^o$ in order to ease comparison with the other results. 
Note that although the data is not well fit by this
model, the results are consistent with those reported by previous experiments.}
\vspace{0.2cm}
\begin{tabular}{ccccc}
Decay Mode &{CLEO II (3 Resonance)}&{E687} &{Mark III} &{E691 (Rotated)} \\ \hline
$K^-\rho^+$   &  $0$ (fixed) & $0$ (fixed) & $0$ (fixed) & $0 \pm 7 $\\
$K^{*-}\pi^+$ & $166.7 \pm 2.0$ & $162 \pm 10$& $154\pm 11$ & $-152 \pm 9 $\\
$\overline{K}^{*0}\pi^0$&$-7.8 \pm 2.2$& $-2\pm 12$& $7\pm 7$& $127 \pm 9 $\\
Non-resonant & $59.7 \pm 2.0$ & $-122 \pm 10$ & $52\pm 9$& $-40 $ (fixed)\\
\end{tabular}
\end{table} 

Since we have at least a factor of ten more statistics for this analysis,
one should not be surprised that more resonances are needed to accurately 
represent the data. The question now becomes how best to determine which 
additional resonances to include.
We have tried two procedures: a) adding all possible resonances and subsequently 
removing those which do not contribute significantly, and b) adding new 
resonances one at a time and choosing the best additional one at each iteration,
stopping when no additional resonances contribute significantly.  Both of these
methods lead us to the same results, hence only the first one is described below.

\begin{figure} 
\centerline{\psfig{file=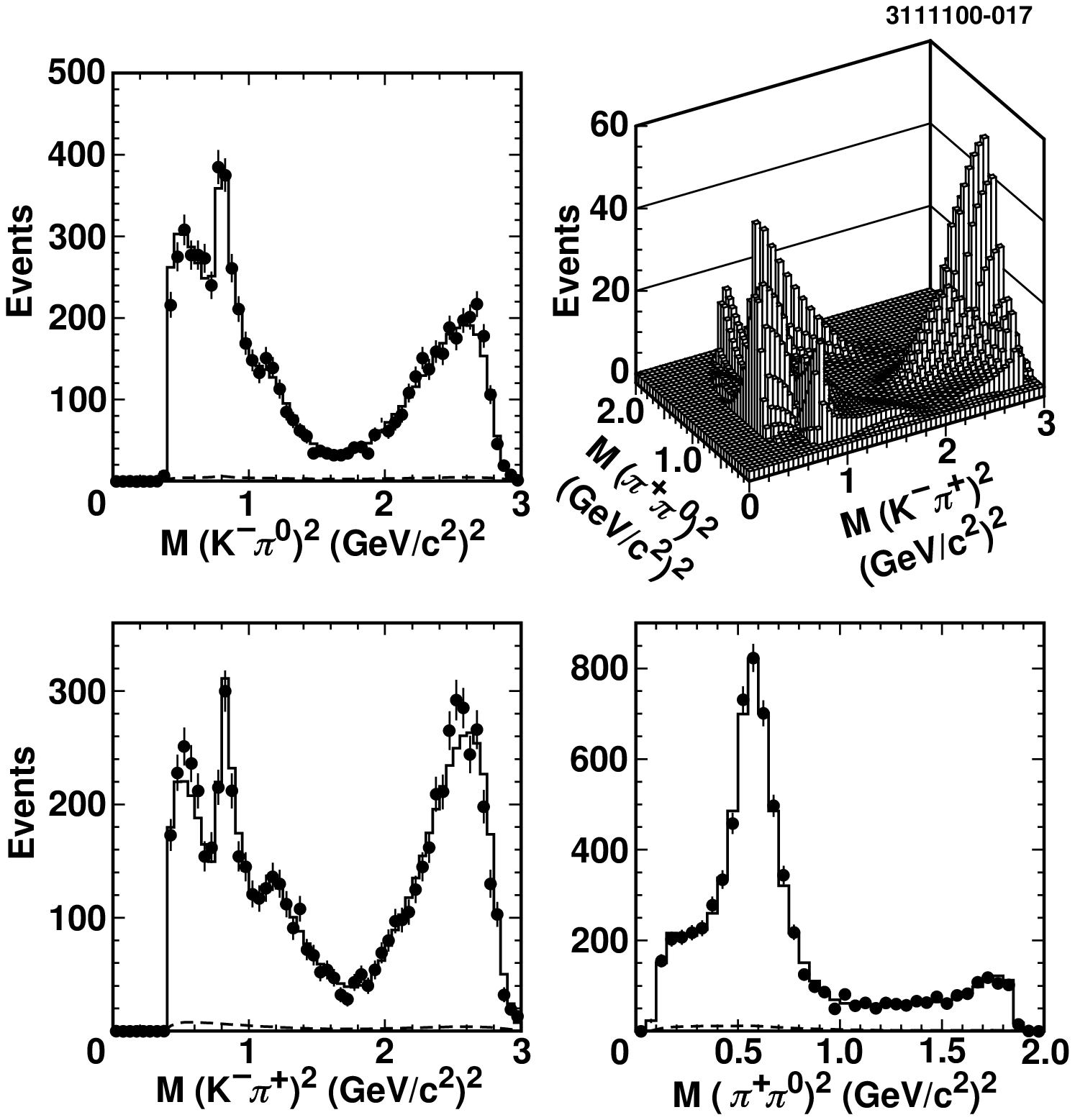,width=6in}}
\vspace{0.5cm}
\caption{\label{fig:s0fit}
The results of fitting the \dkppz ~data with all known resonances
likely to contribute (see Table~\ref{tbl:resonances}).
The efficiency corrected fit is shown as a Manhattan plot (top right), 
and as projections onto the three mass squared variables of both fit 
(histogram) and data (points).
The dashed line shows the level of the background.}
\end{figure}

We begin by fitting the Dalitz Plot with all known 
resonances which can possibly contribute to this decay, as listed
in Table~\ref{tbl:resonances}~\cite{ref:pdg98}.  
The results of this fit are shown in the ``All Resonances'' column of
Table~\ref{tbl:dalsubtract}, and in Figure~\ref{fig:s0fit}. There are 
five resonances which have fit fractions that are less than one standard deviation
away from zero: $\overline{K}^*_3(1780)^0$,
$K^*_3(1780)^-$, $\overline{K}^*(1410)^0$, $K^*(1410)^-$ and
$\overline{K}^*(1680)^0$.  Two other resonances, $\overline{K}^*_2(1430)^0$
and $K^*_2(1430)^-$, have fit fractions close to zero.  When the first five
resonances are removed and the fit repeated, the fit fractions of these last
two resonances do become consistent with zero, and hence are also removed.

\begin{table}
\begin{center}
\caption{\label{tbl:resonances}The resonances considered when fitting the \dkppz ~Dalitz Plot, along
with the masses and widths used when evaluating the matrix element.}
{\normalsize
\vspace{0.2cm}
\begin{tabular}{|l|ccc|}
& \multicolumn{3}{c|}{Parameters}\\ \hline
Resonance                  & $J^P$  & Mass (GeV/c$^2$)    & Width (GeV/c$^2$)  \\ 
\hline
$\rho(770)^+$               &$1^-$  & $0.770\pm 0.001$   & $0.1507\pm 0.0011$ \\
$\overline{K}^*(892)^0$     &$1^-$  & $0.8961\pm 0.0003$  & $0.0505\pm 0.0006$ \\
$K^*(892)^- $               &$1^-$  & $0.8915\pm 0.0003$  & $0.050\pm 0.001$  \\
$K^*(1410)^-$               &$1^-$  & $1.414\pm 0.015$    & $0.232\pm 0.021$   \\ 
$\overline{K}^*(1410)^0 $   &$1^-$  & $1.414\pm 0.015$    & $0.232\pm 0.021$   \\
$K^*_0(1430)^-$             &$0^+$  & $1.412\pm 0.006$    & $0.294\pm 0.023$   \\ 
$\overline{K}^*_0(1430)^0 $ &$0^+$  & $1.412\pm 0.006$    & $0.294\pm 0.023$   \\ 
$K^*_2(1430)^-$             &$2^+$  & $1.425\pm 0.002$    & $0.098\pm 0.003$  \\ 
$\overline{K}^*_2(1430)^0 $ &$2^+$  & $1.432\pm 0.001$    & $0.109\pm 0.005$   \\ 
$\rho(1450)^+$              &$1^-$  & $1.465\pm 0.025$    & $0.310\pm 0.060$   \\
$\rho(1700)^+$              &$1^-$  & $1.700\pm 0.020$    & $0.240\pm 0.060$   \\
$K^*(1680)^-$               &$1^-$  & $1.717\pm 0.027$    & $0.322\pm 0.110$   \\
$\overline{K}^*(1680)^0 $   &$1^-$  & $1.717\pm 0.027$    & $0.322\pm 0.110$   \\
$\overline{K}^*_3(1780)^0$  &$3^-$  & $1.776\pm 0.007$    & $0.159\pm 0.021$   \\
$K^*_3(1780)^- $            &$3^-$  & $1.776\pm 0.007$    & $0.159\pm 0.021$   \\
\end{tabular} }
\end{center}
\end{table} 

\begin{table}
\begin{center}
\caption{\label{tbl:dalsubtract}The parameters from the fits to the \dkppz ~Dalitz 
Plot with all resonances included (``All Resonances'' column), 
and after we remove resonances consistent with zero fit fraction 
(``Final Resonances'' column).  
The $\rho(1450)^+$ and $\rho(1700)^+$  contributions are discussed in the text.}
{\normalsize
\vspace{0.2cm}
\begin{tabular}{|l|cc|cc|}
&\multicolumn{2}{c|}{All Resonances}&\multicolumn{2}{c|}
{Final Resonances} \\ \hline
Component & Phase (degrees)&Fit Fraction (\%)&Phase (degrees)&Fit Fraction (\%) \\ \hline
$\overline{K}_3(1780)^0$ & $263 \pm 16$ & $0.3 \pm 7.5$ & & \\ 
$K_3(1780)^- $ & $86 \pm 12$ & $0.5 \pm 2.9$ & &  \\ 
$\overline{K}^*(1680)^0 $& $175 \pm 25$ & $0.4 \pm 0.5$ & & \\ 
$K^*(1680)^-$& $67 \pm 19$ & $1.0 \pm 0.5$ & $103 \pm 8$ & $1.3 \pm 0.3$ \\
$\rho(1700)^+$ & $149 \pm 8$ & $75 \pm 18$ &$171 \pm 6$&$5.7 \pm 0.8$ \\
$\rho(1450)^+$ & $-45 \pm 10$ & $34 \pm 11$ & &\\ 
Non Res.& $30 \pm 5$ & $9.1 \pm 1.3$ & $31 \pm 4$ & $7.5 \pm 0.9$ \\
$\overline{K}^*(1410)^0 $& $279 \pm 52$ & $0.1 \pm 0.2$  & &\\ 
$\overline{K}_2(1430)^0 $& $148 \pm 13$ & $0.3 \pm 0.14$ & & \\ 
$\overline{K}_0(1430)^0 $& $168 \pm 5$ & $8.0 \pm 1.3$ & $166 \pm 5$&$4.1 \pm 0.6$ \\ 
$K^*(1410)^-$ & $152 \pm 31$ & $0.2 \pm 0.2$ & & \\ 
$K_2(1430)^-$ & $339 \pm 21$ & $0.12 \pm 0.08$ & & \\ 
$K_0(1430)^-$ & $42 \pm 6$ & $5.6 \pm 1.1$ & $55.5 \pm 5.8$ & $3.3 \pm 0.6$ \\ 
$K^*(892)^- $ & $159 \pm 2.6$ & $12.8 \pm 1.8$ &$163 \pm 2.3$ & $16.1 \pm 0.7 $ \\
$\rho(770)^+$ & $0$(fixed) & $74 \pm 4$ & $0$ (fixed) & $78.7 \pm 2.0 $\\
$\overline{K}^*(892)^0 $ & $2.8 \pm 3.2$ & $11.3 \pm 1.5$ & $-0.2 \pm 3.3 $& $ 12.7 \pm 0.9 $\\ \hline
$\chi^2$ &\multicolumn{2}{c|}{203}&\multicolumn{2}{c|}{257}\\
$-2 \ln {\cal L}$&\multicolumn{2}{c|}{6490}&\multicolumn{2}{c|}{6570} \\
C.L.&\multicolumn{2}{c|}{91.3}&\multicolumn{2}{c|}{94.9} \\ 
\hline 
 \end{tabular} }
\end{center}
\end{table} 

Notice that in the ``All Resonances'' column of Table~\ref{tbl:dalsubtract} 
there are two heavy $\rho$ mesons ($\rho(1450)^+$ and $\rho(1700)^+$) 
which have surprisingly large fit fractions. Both have masses which
place their peak outside the Dalitz Plot, but both are wide enough
($310\pm 60$ MeV/c$^2$ and $240\pm 60$ MeV/c$^2$ respectively~\cite{ref:pdg98}) 
that their tails extend well into the
region of interest, making it difficult to distinguish between them.
Since the fitted phases of these $\rho$'s are very close to being $180^o$
apart, their large fit fractions are assumed to be an artifact of the fit's
inability to tell them apart. Supporting this is the additional fact that 
when both resonances are combined, their net contribution to the fit 
fraction is much smaller, $(9 \pm 2)\%$. 
Since the inclusion of both $\rho$ resonances is probably a
misrepresentation of the contents of the Dalitz Plot, only one of these is 
included in all following fits.  We choose the one which gives the best $\chi^2$ and
goodness of fit, the $\rho(1700)^+$, and consider the $\rho(1450)^+$ 
only when evaluating our systematic errors.

\begin{figure} 
\centerline{\psfig{file=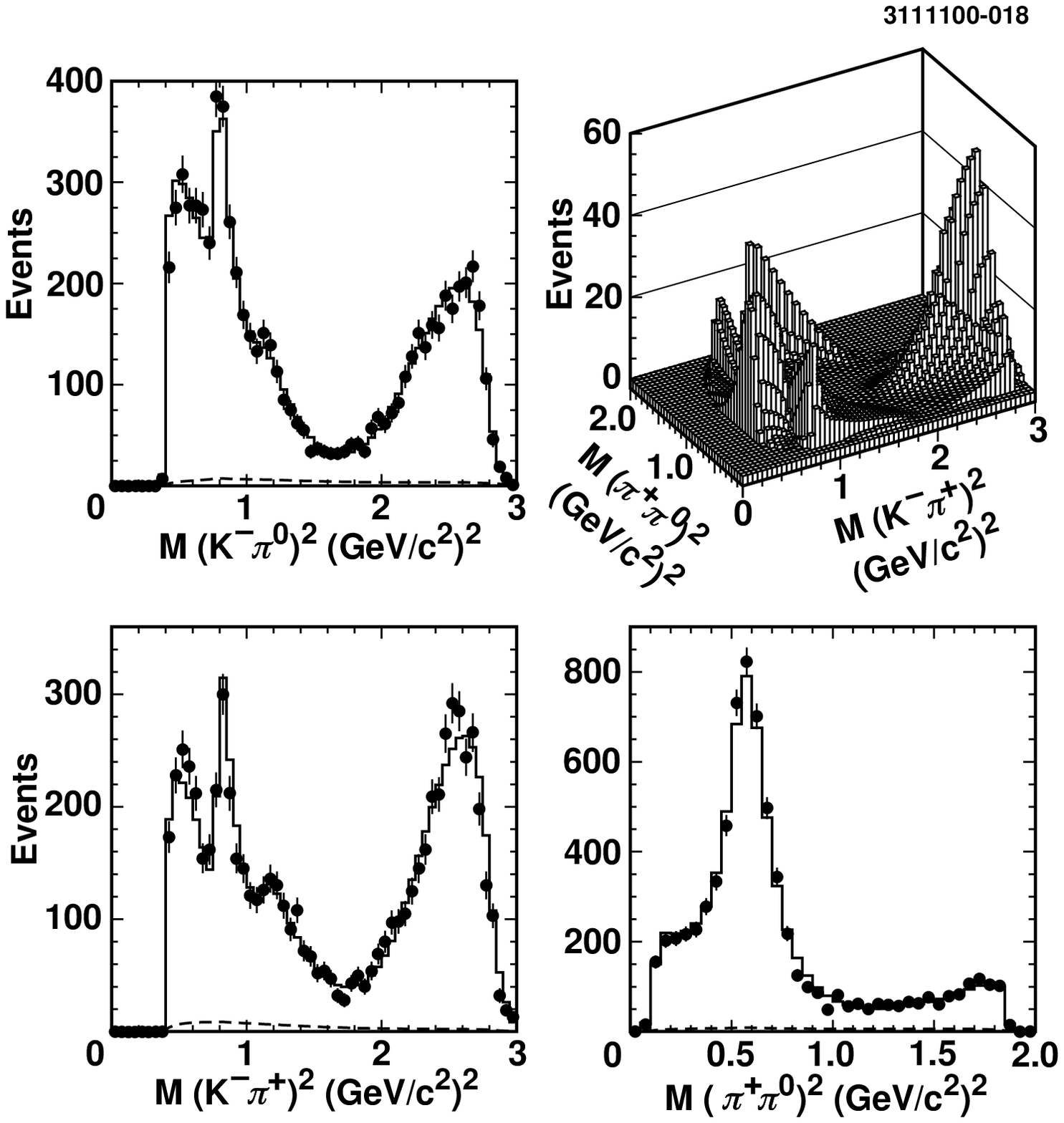,width=6in}}
\vspace{0.5cm}
\caption{\label{fig:finalfit}
The results of fitting the \dkppz ~data with the ``final set'' of components.
The efficiency corrected fit is shown as a Manhattan plot (top right), 
and as projections onto the three mass squared variables of both fit 
(histogram) and data (points).
The dashed line shows the level of the background.}
\end{figure}

After the seven resonances consistent with zero fit fraction are
removed along with the $\rho(1450)^+$ (as discussed above), 
seven resonances remain in addition to the non-resonant component:
$\rho(770)^+$, $K^*(892)^- $, $\overline{K}^*(892)^0$, 
$\rho(1700)^+$, $\overline{K}_0(1430)^0$, $K_0(1430)^-$,
and $K^*(1680)^-$.
Figure~\ref{fig:finalfit} shows the result of fitting the Dalitz 
Plot with these components. The fit fractions and phases are shown in the
``Final Resonances'' column of 
Table~\ref{tbl:dalsubtract}, and the full set of parameters extracted
from this fit is shown in Table~\ref{tbl:finalamps}.

\begin{table}
\begin{center}
\caption{\label{tbl:finalamps} Summary of our best fit to the data with the 
final set of eight components included.}
\vspace{0.1cm}
\hspace{-0.2in}
\begin{tabular}{|lcrlcr||lcrlcr|} 
\multicolumn{6}{|c||}{Signal Parameters} & \multicolumn{6}{c|}{Background Parameters} \\ \hline
\hspace{0.0in}&$a_{nr}$ &\hspace{0.0in}&\hspace{0.0in}&  $ 1.75 \pm 0.12 $&\hspace{0.0in} & & &&& & \\
&$a_{\rho^+}$ &&& $ 1.00 $ (fixed)& &							&$B_0$ &&& $1.0 \pm 0.0$ & \\                  
&$a_{K^{*-}}$ &&& $ 0.44 \pm 0.01 $& &							&$B_x$ &&& $-1.206 \pm 0.001$ & \\                                
&$a_{\overline{K}^{*0}}$ &&& $0.39 \pm 0.01 $&& 					&$B_y$ &&& $-0.74 \pm 0.23$ & \\                            
&$a_{K_0(1430)^-}$ &&& $0.77 \pm 0.08 $&& 						&$B_{x^2}$ &&& $0.468 \pm 0.001$ & \\                            
&$a_{\overline{K}_0(1430)^{0}}$ &&& $0.85 \pm 0.06 $&& 					&$B_{xy}$ &&& $0.842 \pm 0.008$ & \\                        
&$a_{\rho(1700)^+}$ &&& $2.50 \pm 0.19 $&& 						&$B_{y^2}$ &&& $0.168 \pm 0.001$ & \\                           
&$a_{K^*(1680)^-}$ &&& $2.50 \pm 0.3 $&& 					        &$B_{x^3}$ &&& $-0.055 \pm 0.001$ & \\ \cline{1-6}
&$\phi_{NR}$ &&& $ 31.2^o \pm 4.3^o $ && 						&$B_{x^2y}$ &&& $-0.16 \pm 0.06$ & \\                       
&$\phi_{\rho^+}$ &&& $ 0^o $ (fixed)& &							&$B_{xy^2}$ &&& $-0.188 \pm 0.001$ & \\                        
&$\phi_{K^{*-}}$&&& $ 163 \pm 2.3^o$& &							&$B_{y^3}$ &&& $0.077 \pm 0.001$ & \\                      
&$\phi_{\overline{K}^{*0}}$ &&& $ -0.2^o \pm 3.3^o$&& 					&$B_{\ksz}$ &&& $(3.4 \pm 0.1) \times 10^{-5}$ & \\                        
&$\phi_{K_0(1430)^-}$ &&& $55.5^o \pm 5.8^o $&& 					&$B_{\rho}$ &&& $(4.27 \pm 0.05) \times 10^{-4}$ & \\          
&$\phi_{\overline{K}_0(1430)^{0}}$ &&& $ 166^o\pm 5^o $&& 				&$B_{K^{*-}}$ &&& $(9.64 \pm 0.01)\times 10^{-5}$ & \\        
&$\phi_{\rho(1700)^+}$ &&& $171^o \pm 6^o $&& 						&             &&&    & \\ 
&$\phi_{K^*(1680)^-}$ &&& $103^o \pm 8^o $&&                                            &             &&&    & \\
\hline
&Signal Fraction &&& $0.968 \pm 0.007$ &&&&&&& \\
&$-2 \ln {\cal L}$ &&& 6570 &&&&&&& \\
&Conf. Level. &&& 94.9\% & &&&&&&\\
&$\chi^2$ &&& 257 & &&&&&&\\ 
\end{tabular}
\end{center}
\end{table} 

As a curious side note, if a single vector ($K^-\piz$) resonance with a floating
mass and width is added in place of the four new ``standard resonances'' 
discussed above, a good fit can obtained. Unfortunately, while this new 
resonance has a reasonable mass of 1.406 \gev, it prefers a 
negative width of $\Gamma = -0.25$ \gev ~which does not seem to represent 
the underlying dynamics we are trying to measure.
It is possible that the desire for this resonance is an indication of an 
inaccuracy of the formalism used for the resonance shapes, or an indication that
multiple resonances are needed (as we have assumed).
We note that the optimum set of seven resonances used above, all of
which have positive widths, provide a fit which has a lower $\chi^2$
than the inclusion of this single unphysical state.

Other experiments have reported evidence of a light scalar 
($\pi^+\pi^-$) resonance, the $\sigma$, 
in $D^+\rightarrow \pi^-\pi^+\pi^+$ decays~\cite{ref:sigma791}, 
as well as evidence of a scalar ($K^-\pi^+$) resonance, the $\kappa$, in 
$D^+\rightarrow K^-\pi^+\pi^+$ decays ~\cite{ref:kappa791}. 
Since a significant fit fraction for $D^+\rightarrow \kappa\pi^+$ 
has been reported by these authors, we have searched for a scalar 
$\kappa\rightarrow K^-\pi^+$ resonance in the $D^0\rightarrow \kappa\pi^0$ 
channel,  fixing the mass and width of the $\kappa$ to the values reported 
in~\cite{ref:kappa791}, (0.815 \gev~ and 0.560 \gev~ respectively).
We find a fit fraction consistent with zero ($0.4\pm 0.3$\%), 
and see no improvement in the confidence level of the fit with this additional
resonance included.  We have also allowed the mass and width of the $\kappa$ to 
float in the fit, and again see no significant contribution.

Lastly, since this analysis considers only $D^0$ mesons produced from a decaying
$D^{*+}$ in the mode $D^{*+} \to D^0 \pi^+_s$, we have the ability to 
divide our data into separate $D^0$ and $\overline{D^0}$ samples by simply 
considering the sign of the $\pi^\pm_s$ from the $D^{*\pm}$ decay. 
The Dalitz Plots of these samples can then be fitted separately and compared
in a search for CP violation.  We have fitted these samples with the 
same set of resonances described above, and the results are shown in 
Table~\ref{tbl:acp}.  When performing these fits, the efficiency functions
were found separately for the $D^0$ and $\overline{D^0}$ samples, however a common
background shape was assumed. Forming a simple $\chi^2$ between the two sets
of fit parameters we find $\chi^2_{cp} = 16.2$ for 14 degrees of freedom.

We calculate an integrated CP asymmetry across the Dalitz Plot by evaluating 
\begin{equation}
 {\cal A}_{cp} = \int { |{\cal M}_{D^0}|^2 - |{\cal M}_{\overline{D^0}}|^2 \over{
|{\cal M}_{D^0}|^2 + |{\cal M}_{\overline{D^0}}|^2 }} d{\cal DP}
\end{equation}
and obtain ${\cal A}_{cp} = -0.031 \pm 0.086$, consistent with zero. 
Note that this number is not dependent on the number of $D^0$ and $\overline{D^0}$
candidates in our data sample, but rather on the shapes of these 
distributions in the respective Dalitz Plots.

\begin{table}
\begin{center}
\caption{\label{tbl:acp} Fit results when the $D^0$ and $\overline{D^0}$ samples are
considered separately.}
\vspace{0.2cm}
\begin{tabular}{|l|cc|cc|}
&\multicolumn{2}{c|}{$D^0$ Sample}&\multicolumn{2}{c|}{$\overline{D^0}$ Sample} \\ \hline
 Component                &Amplitude     &Phase (degrees)  &Amplitude     &Phase (degrees)  \\ 
 \hline                                  
$\rho(770)^+$             &$1.0\pm 0.0 $    &$0^o$(fixed)     &$1.0\pm 0.0 $ &$0^o$(fixed)  \\
$K^*(892)^- $             &$0.433\pm 0.034$ &$168.9\pm  3.3$ &$0.442\pm 0.015 $ &$157.8\pm  3.4 $ \\
$\overline{K}^*(892)^0 $  &$0.391\pm 0.026$ &$   1.3\pm  3.7$ &$0.410\pm 0.022 $ &$  -4.9\pm  4.9 $ \\ 
$\rho(1700)^+$            &$2.590\pm 0.538$ &$ 175.0\pm  7.5$ &$2.720\pm 0.272 $ &$ 163.9\pm  7.6 $ \\
$\overline{K}_0(1430)^0 $ &$0.989\pm 0.124$ &$ 173.9\pm  8.2$ &$0.774\pm 0.089 $ &$ 159.3\pm  8.1 $ \\ 
$K_0(1430)^-$             &$0.701\pm 0.211$ &$  59.0\pm 10.0$ &$0.917\pm 0.117 $ &$  55.0\pm  7.1 $ \\ 
$K^*(1680)^-$             &$2.567\pm 1.540$ &$ 107.4\pm 69.2$ &$2.060\pm 0.423 $ &$ 106.4\pm 13.5 $ \\
Non Res.                  &$1.840\pm 0.146$ &$  39.9\pm  7.9$ &$1.780\pm 0.160 $ &$  21.3\pm  6.0 $ \\
\hline
$\chi^2$                  &\multicolumn{2}{c|}{227}    &\multicolumn{2}{c|}{233} \\
$-2 \ln {\cal L}$         &\multicolumn{2}{c|}{3237}    &\multicolumn{2}{c|}{3302} \\
C.L.(\%)                  &\multicolumn{2}{c|}{93.1}    &\multicolumn{2}{c|}{80.7} \\ 
\end{tabular} 
\end{center}
\end{table} 

\section{Systematic Uncertainties}
\label{sec:systematics}
After finding the best fit to the data, we must attempt to estimate the
systematic uncertainties on the fit parameters. There are several possible
sources: the background, the efficiency, 
biases due to experimental resolution, and the modeling of the decay. 
These contributions are discussed in
order, and the final systematic errors are shown in Table~\ref{tbl:sys},
where experimental and model dependent sources of systematic uncertainty
are summarized in detail.

The background was modeled by the choice of sideband sample that gave
the best parameterization of the vetoed data sample from Monte
Carlo. Furthermore, the background parameters were allowed to float in
our fits to the data, constrained only by the covariance matrix from
the fit that determined the nominal background function.
To search for any systematic effects due the background parameterization, 
the fitting procedure was repeated for a number of different sideband
choices.  Because the background fraction our sample is a mere 3.3\%, 
or about 230 out of the 7070 events in the Dalitz plot, these
changes have a minimal effect on the fit parameters.  We use the RMS spread 
of these results as our estimate of the systematic error due to our 
choice of background parameterization. 
These values are shown in the ``Bkgnd'' column of Table~\ref{tbl:sys}.

To obtain the efficiency across the
Dalitz Plot, signal Monte Carlo events were fit to a cubic
polynomial. As a check, we have allowed this polynomial to float in
our fit (as was done with the background) subject to a $\chi^2$ constraint
from its covariance matrix in the likelihood function ({\it i.e.} setting $E_{sys}=1$ 
in Equation~\ref{eq:chisquared}). 
If the efficiency is not well modeled by a cubic polynomial, there could
still be an effect that this check would fail to find. To search for
this we tried a local smoothing algorithm rather than the
global polynomial fit. The efficiency was smoothed by fitting either
nine or twenty-five neighbors around each bin with a local plane. 
Each bin's efficiency value was then replaced by the height of this 
plane interpolated to its center. As a final check, we used the raw
measurements of the efficiency in each bin of our fits. We conclude
that the effects of parameterization of the efficiency function over
the Dalitz Plot is not a significant source of concern as most of the
fit parameters vary by less than their one sigma error bars in the
above checks. 

Since we make no requirement on the momentum of the charged tracks,
one might worry that low momentum tracks may be poorly measured and
could affect the Dalitz Plot distribution in a way not well modeled by
our Monte Carlo. To search for such a momentum dependent effect, we fit
the data with the additional requirement that all tracks have a
momentum above 350 MeV/c. 
 
The cuts used to obtain our signal determine the structure of our
efficiency. To assess how well the Monte Carlo reproduces the data
distributions, we varied the cuts used in the analysis and fit
the resulting Dalitz distributions. Each cut was relaxed in turn. The
 cuts on the masses, $M_{D^0}$, $\Delta M$ and $M_{\pi^0}$, were
opened to double the size of the signal region. The minimum energy on the
photons was relaxed to 90 MeV, and the requirement on $X_{D^*}$ was
loosened to 0.5. 

The RMS variation in the fit parameters from each of the tests 
described above was taken as our estimate of the systematic 
uncertainty on the efficiency.
These values are shown in the ``Eff'' column of Table~\ref{tbl:sys}.

A final contribution to the experimental systematic error, presented
in column ``Resol'' of Table~\ref{tbl:sys}, is due to the finite resolution
of the Dalitz Plot variables. As a check, we have included the 
effects of smearing when fitting the data. This was done by measuring the
resolution as a function of position across the Dalitz plot and 
numerically convoluting this with the amplitude at each point when
performing the fit.
Again, the parameters vary by less than the statistical errors on 
the nominal best fit, and their variation from 
the nominal values is taken as an estimate of the systematic uncertainty.

The above three systematic error categories (background, efficiency 
and resolution) are summarized in Table~\ref{tbl:sys}. They are combined 
in quadrature to give the total experimental uncertainty, which is shown
in the ``Total'' column under ``Experiment''. 

\begin{table}
\caption{\label{tbl:sys} A summary of the systematic errors on each fit parameter.  
The first two columns show the results from the best fit and the associated statistical errors.
The next four (three) columns summarize the systematic uncertainties due to experimental (modeling)
sources respectively.  Details are provided in the text.}
\vspace{0.2cm}
\hspace{0.1cm}
\begin{tabular}{|c|cc|cccc|ccc|}
   & \multicolumn{2}{c|}{From Fit} & \multicolumn{4}{c|}{Experimental Errors}  & \multicolumn{3}{c|}{Modeling Errors}\\ 
\hline
       Parameter                        &Value  &Stat Err   &Bkgnd  &Eff    &Resol  &Total  &Shape              &Add       &Total \\ 
\hline
$\overline{K}^*(892)^0$ Fit Frac ($\%$) &12.66  &0.91   &0.17   &0.40   &0.24   &0.50   &1.42               &0.38      &1.47\\
$\overline{K}^*(892)^0$ Phase (deg.)     &-0.20  &3.28   &1.06   &1.62   &1.04   &2.20   &6.99               &0.67      &7.02\\
$\rho(770)^+$ Fit Frac ($\%$)           &78.76  &1.93   &0.52   &1.10   &0.53   &1.33   &4.40               &1.33      &4.60\\
$K^*(892)^-$ Fit Frac ($\%$)            &16.11  &0.69   &0.47   &0.53   &0.18   &0.73   &$^{+2.58}_{-0.48}$ &0.59      &$^{+2.65}_{-0.76}$\\
$K^*(892)^-$ Phase (deg.)                &163.40 &2.32   &0.94   &2.62   &1.30   &3.08   &4.20               &1.09      &4.34\\
$K_0(1430)^-$ Fit Frac ($\%$)           &3.32   &0.64   &0.13   &0.60   &0.40   &0.73   &1.16               &0.40      &1.23\\
$K_0(1430)^-$ Phase (deg.)               &55.52  &5.76   &1.20   &2.76   &1.31   &3.28   &$^{-12.8}_{+3.1}$  &2.85      &$^{+4.2}_{-13.1}$\\
$\overline{K}_0(1430)^0$ Fit Frac ($\%$)&4.05   &0.61   &0.15   &0.66   &0.24   &0.72   &$^{+3.04}_{-0.24}$ &0.39      &$^{+3.06}_{-0.46}$\\
$\overline{K}_0(1430)^0$ Phase (deg.)    &165.90 &5.23   &2.39   &3.83   &0.70   &4.57   &11.4               &3.20      &11.8\\
$\rho(1700)^+$ Fit Frac ($\%$)          &5.65   &0.76   &0.20   &0.43   &0.50   &0.68   &5.71               &0.59      &5.74\\
$\rho(1700)^+$ Phase (deg.)              &170.50 &6.07   &1.90   &3.90   &1.50   &4.59   &$^{-54.7}_{+3.3}$  &5.17      &$^{+6.1}_{-54.9}$\\
$K^*(1680)^-$ Fit Frac ($\%$)           &1.33   &0.33   &0.07   &0.32   &0.11   &0.34   &0.17               &0.32      &0.36\\
$K^*(1680)^-$ Phase (deg.)               &103.20 &7.90   &3.71   &5.91   &2.00   &7.26   &9.21               &9.89      &13.5\\ 
Non Res Fit Frac ($\%$)                 &7.50   &0.95   &0.35   &0.42   &0.05   &0.55   &$^{+5.54}_{-0.79}$ &0.41      &$^{+5.56}_{-0.89}$\\
Non Res Phase (deg.)                     &31.20  &4.28   &1.28   &5.08   &1.70   &5.51   &$^{-14.4}_{+3.5}$  &1.19      &$^{+3.7}_{-14.4}$\\
\end{tabular}
\end{table} 

Modeling systematic errors can arise from our choice of
resonances and the uncertainty in their shapes. 
In Section~\ref{sec:theory} we motivated
our choice of parameterization of the intermediate
resonances; however, other groups have used different functional forms in their
fits~\cite{ref:e687,ref:e691}. We varied these shapes to study
any systematic effects resulting from our choice. We examine three
variations: (i) the Zemach formalism~\cite{ref:zemach1} which enforces the
transversality of the mesons by using $M_{AB}^2$ rather than $M_{r}^2$ 
in the denominator of the spin sums, (ii) a
simple cosine distribution for the spin sum and (iii) a
non-relativistic rather than relativistic Breit-Wigner in the propagator.
Further consideration was also given to the radial parameters used in the form factors, 
which were varied between $0\ {\rm GeV}^{-1}$ and $3\ {\rm GeV}^{-1}$ for the
intermediate resonances and between $0\ {\rm GeV}^{-1}$ and $10\ {\rm
GeV}^{-1}$ for the $D^0$ meson.
The masses and widths of the intermediate resonances were 
allowed to vary within the known errors~\cite{ref:pdg98}.  
The non-resonant contribution was described in our fits by a constant
term, but as a check we also modeled it by a linear function  or a shape given by
the spin structure without the Breit-Wigner amplitudes~\cite{ref:nr1}. 

The above tests were used to explore the
systematic dependance of the fit parameters on the way the physics
was modeled. 
The variations using a simple cosine 
distribution in place of the spin sum and using a spin structured 
rather than constant non-resonant component resulted in fits with 
significantly worsened $\chi^2$ (368 and 322 respectively), and 
are not considered when assigning a systematic error as the data
suggests these forms could not be correct. 
We take the largest of the remaining variations as the systematic 
error due to our choice of modeling shapes, and the results are shown
in the ``Shape'' column of Table~\ref{tbl:sys}.

The final systematic check is on our choice of which resonances to
include.  For example, there is only a slight preference for the
$\rho(1700)^+$ over the $\rho(1450)^+$ based on the goodness of fit.
To account for this uncertainty, both fits were performed and the
variation of the parameters were noted. Fits were also performed
which included additional resonances from Table~\ref{tbl:resonances}.  
The RMS variation in the fit parameters from the above checks is 
presented in the ``Add'' column of Table~\ref{tbl:sys}. 

We also considered the effects of removing resonances, and two of these
studies deserve further comment. The first is the removal of the $K^*(1680)^-$. 
We considered this because the final fit fraction for this resonance 
is a rather small $1.3\pm 0.3$\%.
When the $K^*(1680)^-$ is removed the $\chi^2$ increases from 257 to 316
indicating that this resonance should remain. The parameters for this fit are 
shown in the ``Removed $K^*(1680)^-$'' column of Table~\ref{tbl:minus}.
For comparison, when the other ``new'' resonances, $\overline{K}_0(1430)^0$, 
$K_0(1430)^-$, and $\rho(1700)^+$, are removed, the $\chi^2$ increases 
to 379, 348, and 381 respectively. 
The second case which deserves special attention is the removal of the non-resonant

component.  Some theoretical models, such as chiral perturbation
theory~\cite{ref:nonr},  prefer a small non-resonant
component, suggesting it proceeds only by the coherent
sum of two body decays. When this test is performed on our data,
the resulting $\chi^2$ jumps to 411, suggesting that a non resonant 
component is indeed present. The parameters for this fit are 
shown in the ``Removed Non Resonant'' column of Table~\ref{tbl:minus}.

Since removal of any of the fit components causes a significant increase
in the $\chi^2$ of the fit, these variations were not included in the modeling
systematic error.  To obtain the final model dependent systematic error we add 
the ``Shape'' and ``Add'' columns of Table~\ref{tbl:sys} 
in quadrature to obtain the result shown in the ``Total'' column
under ``Model''.

\begin{table}
\begin{center}
\caption{\label{tbl:minus} Fit results after removal of the either the 
$K^*(1680)^-$ resonance or the non-resonant component.  
See Section~\ref{sec:systematics} for discussion.}
\vspace{0.2cm}
\begin{tabular}{|l|cc|cc|}
&\multicolumn{2}{c|}{Removed $K^*(1680)^-$}&\multicolumn{2}{c|}{Removed Non Resonant} \\ \hline
Component                 &Phase (degrees)     &Fit Fraction (\%)   &Phase (degrees)  &Fit Fraction  (\%) \\ 
\hline
$\rho(770)^+$             &$0$ (fixed)     &$80.8\pm 8.5$  &$0$(fixed)      &$77.8\pm 1.8$ \\
$K^*(892)^- $             &$157\pm 6.7$  &$13.8\pm 1.0$  &$161\pm 2.2$  &$18.2\pm 0.7$ \\
$\overline{K}^*(892)^0 $  &$-4.7\pm 5.7$ &$14.5\pm 1.3$  &$-1.5\pm 2.8$ &$10.7\pm 0.8$ \\ 
$\rho(1700)^+$            &$161\pm 20$   &$6.7\pm 0.8$   &$161\pm 5$    &$5.4\pm 0.8$  \\
$\overline{K}_0(1430)^0 $ &$164\pm 9$    &$4.4\pm 0.5$   &$194\pm 9$    &$1.0\pm 0.3$  \\ 
$K_0(1430)^-$             &$47.8\pm 3.6$ &$4.5\pm 0.7$   &$11\pm 4$     &$5.2\pm 0.7$  \\ 
$K^*(1680)^-$             &$0$             &$0.0$          &$90\pm 5$     &$1.9\pm 0.5$  \\
Non Res.                  &$37\pm 6$     &$7.7\pm 2.6$   &$0$             &$0.0$         \\
\hline
$\chi^2$                  &\multicolumn{2}{c|}{316 }        &\multicolumn{2}{c|}{411 } \\
$-2 \ln {\cal L}$         &\multicolumn{2}{c|}{6653}        &\multicolumn{2}{c|}{6798} \\
C.L.(\%)                  &\multicolumn{2}{c|}{98.5}        &\multicolumn{2}{c|}{0.7 } \\ 
\end{tabular} 
\end{center}
\end{table} 

\section{Summary of Results}

We have fit the distribution of data in the \dkppz ~Dalitz Plot
obtained with the CLEO~II experiment to a coherent sum of seven 
intermediate resonances plus a non-resonant component.  All resonances
are either scalar or vector; no significant tensor contribution was 
found. The non-resonant contribution is significant, and cannot be removed 
without seriously compromising the quality of the fit.  
We see no evidence of a scalar $\kappa\rightarrow K^-\pi^+$ resonance 
in the mass range recently reported by other groups.

The final fit fraction and phase for each component is given in 
Table~\ref{tbl:conclusion}. These fit fractions, multiplied by the world 
average \dkppz ~branching ratio of $(13.9 \pm 0.9)$\% \cite{ref:pdg2000}, 
yield the partial branching fractions shown in Table~\ref{tbl:fractions}.
The error on the world average branching ratio is incorporated by 
adding it in quadrature with the experimental systematic errors on the 
fit fractions to give the experimental systematic error on the 
partial branching fractions.
Note that due to interference the fit fractions do not add to unity, 
and consequently the partial branching fractions do not sum to 
the world average.

By separately fitting the \dkppz ~and \dkppzbar ~Dalitz Plots, 
we have calculated the integrated CP asymmetry across the Dalitz Plot
to be ${\cal A}_{cp} = -0.031 \pm 0.086$.

\begin{table} 
\caption{\label{tbl:conclusion} Final fit results. The errors shown are statistical, 
experimental systematic, and modeling systematic respectively, as discussed in 
Section~\ref{sec:systematics} and summarized in Table~\ref{tbl:sys}. }
\vspace{0.2cm}
\begin{tabular}{|l|cc|} 
  Mode                          & Fit Fraction                                 & Phase  (degrees)     \\ 
\hline
$\rho(770)^+ K^-$               &$0.788\pm 0.019\pm 0.013 \pm0.046$            & $0.0$ (fixed)                          \\
$K^*(892)^- \pip$               &$0.161\pm 0.007\pm 0.007 ^{+0.026}_{-0.008}$  & $163 \pm 2.3 \pm 3.1 \pm 4.3$          \\
$\overline{K}^*(892)^0 \piz$    &$0.127\pm 0.009\pm 0.005 \pm0.015$            & $-0.2 \pm 3.3 \pm 2.2 \pm 7.0$         \\
$\rho(1700)^+ K^-$              &$0.057\pm 0.008\pm 0.007\pm 0.006$            & $171 \pm 6 \pm {5} ^{+6.1}_{-55}$      \\
$\overline{K}^*_0(1430)^0 \piz$ &$0.041\pm 0.006\pm 0.007^{+0.031}_{-0.005}$   & $166 \pm 5 \pm {4.6} \pm 12$           \\
$K^*_0(1430)^- \pip$            &$0.033\pm 0.006\pm 0.007\pm 0.012$            & $55.5 \pm 5.8 \pm {3.3} ^{+4.2}_{-13}$ \\
$K^*(1680)^- \pip$              &$0.013\pm 0.003\pm 0.003\pm 0.003$            & $103 \pm 8 \pm {7} \pm 14$             \\
Non Resonant                    &$0.075\pm 0.009\pm 0.006 ^{+0.056}_{-0.009}$  & $31 \pm 4 \pm {5.5} ^{+14}_{-3.7}$     \\
\end{tabular}
\end{table} 

\begin{table} 
\caption{\label{tbl:fractions}   
Partial branching fractions calculated by combining our fit fractions 
with the previously measured \dkppz~branching ratio as described in the text.
The errors shown are statistical, experimental systematic, and modeling 
systematic respectively.}
\vspace{0.2cm}
\begin{tabular}{|l|c|} 
Mode                            & Partial Branching Fraction   \\ 
\hline
$B(D^0\to\rho(770)^+ K^-)\times B(\rho(770)^+\to\pi^+\pi^0)$                           & $0.109 \pm 0.003  \pm 0.007  \pm 0.006$          \\
$B(D^0\to K^*(892)^- \pip)\times B(K^*(892)^-\to K^-\pi^0)$                            & $0.022 \pm 0.001  \pm 0.002  ^{+0.004}_{-0.001}$ \\
$B(D^0\to\overline{K}^*(892)^0 \piz)\times B(\overline{K}^*(892)^0\to K^-\pi^+)$       & $0.018 \pm 0.001  \pm 0.001  \pm 0.002$          \\
$B(D^0\to\rho(1700)^+ K^-)\times B(\rho(1700)^+\to\pi^+\pi^0)$                         & $0.008 \pm 0.001  \pm 0.001  \pm 0.001$          \\
$B(D^0\to\overline{K}^*_0(1430)^0 \piz)\times B(\overline{K}^*_0(1430)^0\to K^-\pi^+)$ & $0.006 \pm 0.001  \pm 0.001  ^{+0.004}_{-0.001}$ \\
$B(D^0\to K^*_0(1430)^- \pip)\times B(K^*_0(1430)^-\to K^-\pi^0)$                      & $0.005 \pm 0.001  \pm 0.001  \pm 0.002$          \\
$B(D^0\to K^*(1680)^- \pip)\times B(K^*(1680)^-\to K^-\pi^0)$                          & $0.0018\pm 0.0004 \pm 0.0004 \pm 0.0004$         \\
$B(D^0\to K^-\pi^+\pi^0)$ Non Resonant                                                 & $0.010 \pm 0.001  \pm 0.001  ^{+0.008}_{-0.001}$ \\
\end{tabular}
\end{table} 

We gratefully acknowledge the effort of the CESR staff in providing us with
excellent luminosity and running conditions.
M. Selen thanks the PFF program of the NSF and the Research Corporation, 
A.H. Mahmood thanks the Texas Advanced Research Program,
F. Blanc thanks the Swiss National Science Foundation, 
and E. von Toerne thanks the Alexander von Humboldt Stiftung for support.
This work was supported by the National Science Foundation, the
U.S. Department of Energy, and the Natural Sciences and Engineering Research 
Council of Canada.

\end{document}